\documentclass[12pt]{article}

\usepackage[english]{babel}

\usepackage[a4paper,top=2cm,bottom=2cm,left=3cm,right=3cm,marginparwidth=1.75cm]{geometry}

\usepackage{amsmath}
\usepackage{graphicx}
\usepackage{yfonts}
\usepackage{amssymb}
\usepackage{amsthm,bm}
\usepackage{comment}
\usepackage{bbm}
\usepackage{mathtools}
\newtheorem{thm}{Theorem}[section]
\newtheorem{co}[thm]{Corollary}
\newtheorem{lem}[thm]{Lemma}
\newtheorem{conj}[thm]{Conjecture}
\newtheorem{assumption}[thm]{Assumption}

\newtheorem{pr}[thm]{Proposition}

\newtheorem{definition}[thm]{Definition}

\newtheorem{example}[thm]{Example}

\newtheorem{remark}[thm]{Remark}

\usepackage[colorlinks=true, allcolors=blue]{hyperref}

\newcommand{\E}{\mathbb{E}}

\newenvironment{keywords}
  {\par\noindent\textbf{Keywords: }}
  {\par}

\title{Second-Order Schalkwijk-Kailath Coding for Autoregressive Gaussian Channels}
\author{\footnotesize \centering\begin{tabular}{ccc}
Jun Su & Guangyue Han & Shlomo Shamai (Shitz)\\
The University of Hong Kong &The University of Hong Kong& Technion-Israel Institute of Technology\\
 email:  junsu@hku.hk& email:  ghan@hku.hk& email: sshlomo@ee.technion.ac.il\\
\end{tabular}}
\begin{document}
\maketitle
\renewcommand{\thefootnote}{\fnsymbol{footnote}}
\footnotetext{\hspace*{0mm} A preliminary version of this work has been presented in IEEE International Symposium on Information Theory (ISIT) 2026, DOI: 10.1109/ISIT62367.2026.11653872.}

\begin{abstract}
We study noiseless-feedback communication over Gaussian channels with stationary autoregressive (AR) noise of arbitrary finite order. We introduce a class of Gaussian linear-feedback coding schemes based on the celebrated Schalkwijk--Kailath (SK) coding scheme, called SK(2), whose message process follows a second-order deterministic recursion, and derive a closed-form characterization of its maximal achievable rate under average power constraint $P$. As a first-order benchmark, we formulate SK(1) as a sign-consistent, branch-complete version of the real constant signed-ratio specialization of Butman's linear-feedback construction. For stationary AR($p$) Gaussian channels, optimizing over recursions with both characteristic roots outside the unit circle gives the same rate supremum as optimizing over all SK(2) recursions and yields a rate no smaller than the maximal SK(1) rate. The SK(2) scheme achieves feedback capacity for the AWGN channel, stationary AR(1) Gaussian channels, and a subclass of stationary AR(2) Gaussian channels obtained by interleaving two independent AR(1) noise processes. For the interleaved subclass, we show that a genuine SK(2) scheme strictly outperforms SK(1). Numerical comparisons illustrate further strict rate gains over a broader range of AR(2) parameters. These results show that the SK(1) scheme is not universally optimal beyond AR(1) noise and disprove a corrected form of Butman's conjecture.
\end{abstract}


\section{Introduction} \label{sec:introduction}
We consider the following additive colored Gaussian noise (ACGN) channel with noiseless feedback:
\begin{equation}\label{Channel-ACGN}
   Y_i = X_i(M,Y_1^{i-1}) + Z_i,\quad i=1,2,\dots,
\end{equation}
where $M$ denotes the transmitted message, and $\{Z_i\}_{i=1}^\infty$ is a zero-mean stationary Gaussian noise process independent of $M$. The channel input $X_i$ at time $i$ may causally depend on the message $M$ and the previous channel outputs $Y_1^{i-1}\triangleq\{Y_1,\ldots,Y_{i-1}  \}$. For each block length $n$, an admissible feedback code satisfies the average power constraint $P>0$, namely,
$$
   \frac{1}{n}\sum_{i=1}^n\E[X^2_i(M,Y_1^{i-1})] \le P.
$$
Let $C_{\mathrm{FB}}(P)$ denote the capacity of the ACGN channel \eqref{Channel-ACGN}, commonly referred to as the {\em Gaussian feedback capacity}.

It is well-known that, without feedback, the capacity of the ACGN channel \eqref{Channel-ACGN}, denoted by $C(P)$, can be determined by the water-filling method \cite{Shannon1949}. In the special case where the noise process is white, i.e., $Z_i=W_i$ with $\{W_i\}$ a white Gaussian process with zero mean and unit variance, the channel reduces to the additive white Gaussian noise (AWGN) channel, whose capacity was first determined by Shannon \cite{Shannon1949,shannon1948mathematical}; see also \cite{wolfowitz1964,wolfowitz1968note}. More precisely, Shannon \cite{shannon1956zero} proved that for this channel, feedback does not increase capacity, that is, the feedback capacity $C_{\mathrm{FB}}(P)$ coincides with the nonfeedback capacity $C(P)$; see also \cite{kadota1971capacity,kadota1971mutual}. Nevertheless, feedback can greatly simplify coding and improve reliability. A significant later result, due to Schalkwijk and Kailath \cite{schalkwijk1966coding2, schalkwijk1966coding}, showed that a simple linear feedback coding scheme, commonly referred to as the SK coding scheme, achieves the capacity of the AWGN channel.

For the ACGN channel \eqref{Channel-ACGN} when the noise process is not white, feedback may increase capacity, and the optimal coding structure is substantially more delicate. Butman \cite{butman1969} formulated linear feedback communication systems for general ACGN channels and derived the corresponding optimal linear signaling structure. For stationary first-order autoregressive, or AR(1), Gaussian channels under average power constraint \(P\), Butman's equal-energy scheme gives a closed-form achievable rate. Wolfowitz \cite{wolfowitz1975signalling} provided an alternative derivation of this rate and proved optimality within a class of linear signaling schemes. Tiernan and Schalkwijk \cite{Tiernan1974} established upper bounds on $C_\mathrm{FB}(P)$ for AR(1) Gaussian channels with noiseless feedback, while Ozarow \cite{ozarow1990upper,ozarow1990random} studied upper bounds on $C_\mathrm{FB}(P)$ and random coding for ACGN channels with feedback. In particular, Ozarow \cite{ozarow1990random} showed that feedback strictly increases capacity for nonwhite stationary Gaussian noise, while traditional random coding is not generally capacity-achieving in the feedback setting.

Butman subsequently extended the equal-energy construction to general $p$-th-order autoregressive, or AR($p$), Gaussian channels and conjectured that the resulting lower bound is tight \cite{butman1976linear}. As we explain in Section~\ref{subsec:relation_to_butman}, however, the AR($p$) formula in \cite{butman1976linear} contains a sign issue: the transition from the absolute-value recursion to the final polynomial expression requires additional sign-compatibility and branch conditions. Section~\ref{subsec:relation_to_butman} addresses these sign and branch issues and formulates a corrected constant signed-ratio benchmark.

Several later works studied feedback capacity for the ACGN channel \eqref{Channel-ACGN} from dynamic-programming, state-space, and variational viewpoints. Yang et al. \cite{yang2005feedback,yang2007feedback} developed dynamic-programming formulations for finite-state machine channels and power-constrained Gaussian channels with memory. For the first-order moving-average, or MA(1), Gaussian channel, Kim \cite{Kim2006} showed that a modified SK coding scheme achieves the feedback capacity and obtained a closed-form expression for $C_\mathrm{FB}(P)$. Subsequently, Kim \cite{kim2009feedback} gave a variational characterization of the feedback capacity of stationary Gaussian channels. Related refinements and extensions can be found in \cite{liu2018feedback,sabag2023,su2024feedback}. In particular, the first-order autoregressive moving-average, or ARMA(1,1), Gaussian channel is now well understood. Kim \cite[Theorem 6.1]{kim2009feedback} also stated that a $k$-dimensional variant of the SK coding scheme achieves feedback capacity for any ARMA noise of order $k$. However, Derpich and \O stergaard \cite{derpich2022comments} identified a gap in the component-removal argument underlying Kim's general coding result \cite[Corollary 4.4]{kim2009feedback}. A recent preprint by Fay and Sabag \cite{fay2026feedback} reports component removal and a generalized SK construction from finite-dimensional capacity optimizers for nonwhite rational noise spectra. Su, Xu, and Han \cite{su2026hardy} establish the filter-only capacity formula at the level of suprema under the Paley--Wiener condition, without asserting attainment by a single filter.

In this paper, we focus on the stationary AR($p$) Gaussian channel
\begin{equation}\label{channel-arp}
Y_i = X_i + Z_i, \quad i\ge 1,
\end{equation}
where the noise process $\{Z_i\}$ is a real-valued AR($p$) Gaussian process. Its power spectral density \cite{liu2018feedback} is
\begin{equation}\label{sdf-arp}
S_Z(e^{i\theta}) = \vert H_Z(e^{i\theta}) \vert^2 = \frac{1}{\vert L_Z(e^{i\theta}) \vert^2} = \frac{1}{\vert\prod_{k=1}^{p}(1+\beta_ke^{i\theta})\vert^2},
\end{equation}
where the polynomial $L_Z$ has real coefficients, $|\beta_k|<1$ for every $k$, and the driving white Gaussian noise has unit variance. We allow $p=0$, with $L_Z\equiv 1$. 

Unless otherwise stated, all coding results below concern the real stationary AR($p$) Gaussian channel \eqref{channel-arp}--\eqref{sdf-arp} under the average power constraint $P > 0$ stated above. General variational, algebraic, and finite-dimensional convex characterizations of Gaussian feedback capacity are available \cite{kim2009feedback,liu2018feedback,sabag2023}. A central analytical problem is to obtain tractable capacity formulas and optimal coding parameters directly from the noise coefficients. Here, we study this problem through scalar deterministic message recursions and derive explicit rate and power formulas for the second-order class.

We first formulate the SK(1) coding scheme, which provides a branch-complete and sign-consistent formulation of the real constant signed-ratio specialization of Butman's linear-feedback construction \cite{butman1976linear}. Motivated by this formulation, we state a corrected version of Butman's conjecture as the optimality of the SK(1) coding scheme:
$$
    C_{\mathrm{FB}}(P)=C_{\mathrm{SK1}}(P),
$$
where $C_{\mathrm{SK1}}(P)$ is the maximal achievable rate of the SK(1) coding scheme. The main contribution of this work is a closed-form characterization of the information-rate supremum for a Gaussian feedback coding scheme, termed the SK(2) coding scheme, over the stationary AR($p$) channel (\ref{channel-arp}). This rate, denoted by $C_\mathrm{SK2}(P)$, provides a lower bound on $C_\mathrm{FB}(P)$. For certain AR(2) Gaussian channels, we show that the SK(2) coding scheme can strictly outperform the SK(1) coding scheme and hence
$$
    C_{\mathrm{FB}}(P)\ge C_{\mathrm{SK2}}(P)>C_{\mathrm{SK1}}(P),
$$
which disproves the corrected form of Butman's conjecture.

The remainder of the paper is organized as follows. In Section~\ref{sec:the_classical_schalkwijk_kailath_coding}, we review the classical SK coding scheme. In Section~\ref{subsection:the_sk_1_coding_scheme}, we introduce the SK(1) coding scheme and derive its maximal achievable rate in Theorem~\ref{ISK1-thm}. We then clarify its relation to Butman's equal-energy scheme and reformulate Butman's conjecture in terms of \(C_{\mathrm{SK1}}(P)\) in Section~\ref{subsec:relation_to_butman}. Section~\ref{sec:sk2-coding-scheme} contains the main contributions:
\begin{itemize}
\item[$\bullet$] We propose the SK(2) coding scheme for stationary AR($p$) Gaussian channels and derive a closed-form characterization of the maximal information rate within its genuine second-order class (Theorem~\ref{main-theorem-sk2}). A reduction of the remaining root configurations shows that the genuine supremum gives the full SK(2) capacity and is no smaller than the SK(1) capacity (Lemma~\ref{lemma2:label} and Corollary~\ref{main-col-sk2:label}). We further identify a sufficient condition under which a repeated-root scheme achieves twice the SK(1) capacity (Corollary~\ref{co-twice-sk2:label}).


\item[$\bullet$] We identify three capacity-achieving cases: AWGN channels (Corollary~\ref{co-awgn-sk2:label}), stationary AR(1) channels (Proposition~\ref{example-ar1:label}), and an interleaved AR(1) subclass of stationary AR(2) channels (Proposition~\ref{prop:interleaved-AR1}). For the interleaved subclass, a genuinely second-order scheme attains feedback capacity and strictly outperforms SK(1), providing an analytical counterexample to the corrected form of Butman's conjecture (Conjecture~\ref{Butman-conj:label}).
\item[$\bullet$] Numerical examples for more general stationary AR(2) channels illustrate strict improvements of SK(2) over SK(1) across a range of parameters (Example~\ref{example-ar2:label}).
\end{itemize}

\section{The SK(1) Coding Scheme}
It is well known that the SK coding scheme achieves feedback capacity for the AWGN channel under average power constraint $P$. In this section, we extract its core structure into a first-order deterministic message recursion, termed the SK(1) coding scheme. We first recall the classical SK coding scheme as motivation. We then define the SK(1) coding scheme for stationary AR($p$) Gaussian channels \eqref{channel-arp} and derive its maximal achievable rate. Finally, we relate the SK(1) coding scheme to Butman's equal-energy linear-feedback coding scheme \cite{butman1976linear}, point out some technical issues in Butman's treatment, and reformulate Butman's conjecture in terms of the corrected constant signed-ratio lower bound.

\subsection{The Schalkwijk--Kailath Scheme}\label{sec:the_classical_schalkwijk_kailath_coding}
In the spirit of the exposition in \cite[Section 17.1.1]{el2011network}, the classical SK coding scheme can be reformulated as follows. Consider the discrete-time AWGN channel
\begin{equation}\label{AWGN-channel}
Y_i=X_i+W_i,\qquad i \ge 0,
\end{equation}
where $\{W_i\}$ is a white Gaussian process with zero mean and unit variance and $\{X_i\}$ is subject to the average power constraint $P$, i.e., 
$$
\frac{1}{n}\sum_{i=0}^{n-1}\E[X_i^2]\le  P.
$$
The encoding scheme proceeds as follows. At time $i=0$,  the encoder transmits a symbol $X_0$ chosen arbitrarily from the interval $[-\sqrt{P},\sqrt{P}]$,  yielding the output $Y_0 = X_0+W_0$. Because of noiseless feedback, the encoder learns $Y_0$ and can compute the noise realization $W_0 = Y_0 - X_0$. At time $i=1$, the encoder transmits $X_1 = \gamma_1W_0$, where $\gamma_1=\sqrt{P}$ ensures that $\E[|X_1|^2] =P$. For each subsequent time $i\ge 2$, the encoder computes the minimum mean-square error (MMSE) estimate $\E[W_0\mid Y_1^{i-1}]$ based on all previous channel outputs $Y_1^{i-1}$ available via feedback. It then transmits 
\begin{equation}\label{SK-coding-rule}
X_i = \gamma_i (W_0 - \E[W_0\mid Y_1^{i-1}]),
\end{equation}
where $\gamma_i$ is chosen such that $\E[|X_i|^2] = P$ for all $i$. The SK coding scheme, given in (\ref{SK-coding-rule}), admits a recursive interpretation in which the transmitter repeatedly refines the receiver's knowledge of the initial noise $W_0$, or equivalently, the original message $X_0$, across successive uses of the channel. Specifically, the encoding rule can be written as
\begin{equation*}
\begin{aligned}
X_i =  \gamma_i (W_0 - \E[W_0\mid Y_1^{i-1}])= \frac{\gamma_i}{\gamma_{i-1}} \left(X_{i-1} -\E[X_{i-1}\mid Y_1^{i-1}]\right)
\end{aligned}
\end{equation*}
so that each transmission corrects the receiver's estimation error incurred in the previous transmission. 

Furthermore, define the process $V_i \triangleq \gamma_i W_0$. Since $W_0$ is independent of the forward noise $\{W_i\}_{i\ge 1}$, $\{V_i\}$ may be regarded as the underlying {\it message process}. Under this representation, the SK rule takes the following form:
\begin{equation}\label{SK-coding-rule-X}
X_i = V_i - \E[V_i\mid Y_1^{i-1}],
\end{equation}
where $\{V_i\}$ follows the deterministic recursion:
\begin{equation}\label{SK-coding-rule-V}
V_i = \frac{\gamma_i}{\gamma_{i-1}} V_{i-1},\qquad i\ge 2,
\end{equation}
with initial condition $V_1=\sqrt{P}W_0$. With the coefficients chosen to satisfy $\E[X_i^2]=P$ at each refinement step, this SK coding scheme (\ref{SK-coding-rule}), or equivalently, the recursive form (\ref{SK-coding-rule-X}) and (\ref{SK-coding-rule-V}), achieves the feedback capacity of the AWGN channel \cite{schalkwijk1966coding2,schalkwijk1966coding,el2011network}.

The SK coding scheme described above is designed for the AWGN channel \eqref{AWGN-channel}, where the noise samples are uncorrelated. For the general ACGN channel \eqref{Channel-ACGN}, the above recursive representation motivates a natural first-order generalization. We now formalize this generalization as the SK(1) coding scheme. Its relation to Butman's equal-energy coding scheme will be discussed after the corresponding maximal achievable rate characterization in Theorem~\ref{ISK1-thm}.

\subsection{The SK(1) Coding Scheme}\label{subsection:the_sk_1_coding_scheme}
We now introduce a first-order specialization of the SK coding scheme, which we term the SK(1) coding scheme. The name reflects that the underlying message process follows a first-order deterministic recursion. It is obtained from the recursive form in (\ref{SK-coding-rule-X}) and (\ref{SK-coding-rule-V}) by imposing the simple ratio condition
\begin{equation}\label{sk1-ratio}
\frac{\gamma_i}{\gamma_{i-1}}=\gamma\quad\mbox{for all}~i,
\end{equation}
where $|\gamma|>1$. Under this condition, the message process $V=\{V_i\}$ follows a first-order deterministic recursion. Formally, the SK(1) coding scheme is defined by
\begin{equation}\label{SK1-coding-X}
X_i = V_i - \E[V_i\mid Y_1^{i-1}]
\end{equation}
with the message process $\{V_i\}$ such that 
\begin{equation}\label{SK1-coding-message}
    V_1=U,\qquad V_{i+1} = \gamma V_i, \qquad i \ge 1,
\end{equation}
where $U$ is a standard Gaussian random variable, independent of the noise process $Z=\{Z_i\}$. We define the {\em SK(1) information rate capacity}, or simply, {\em SK(1) capacity} under average power constraint \(P\) as
\begin{equation}\label{feedback-capacity-SK1}
C_{\mathrm{SK1}}(P) = \sup_{(V,X)} \bar{I}(V;Y),
\end{equation}
where $\bar{I}(V;Y)$ denotes the mutual information rate between the processes $\{V_i\}$ and \(\{Y_i\}\). The supremum is taken over all pairs \((V,X)\) satisfying \eqref{SK1-coding-X}, \eqref{SK1-coding-message} and the asymptotic average power
constraint
\begin{equation}\label{power-constraint-sk1}
   \varlimsup_{n\to\infty}\frac{1}{n}\sum_{i=1}^n\E[X_i^2]\le P.
\end{equation}
The finite-dimensional Gaussian block channels induced by the linear-feedback constructions considered here are information stable. Standard Gaussian coding arguments therefore make every rate below the limiting mutual information rate reliably achievable under the same asymptotic average-power constraint \eqref{power-constraint-sk1} (see, e.g., \cite[Section III]{kim2009feedback}). Thus, \(C_{\mathrm{SK1}}(P)\) is the maximal achievable information rate within the SK(1) coding class. In particular,
$$
    C_{\mathrm{SK1}}(P)\le C_{\mathrm{FB}}(P).
$$
\begin{remark}
Note that the SK(1) coding scheme, which can be directly transformed into the original SK coding, has been established as capacity-achieving for stationary ARMA(1,1) Gaussian channels with feedback \cite{kim2009feedback}.
\end{remark}

We begin by stating a key lemma, the proof of which is deferred to Appendix \ref{sec:proof_of_lemma_ref_lemma1}. This lemma characterizes the quadratic form associated with the inverse of a structured matrix.
\begin{lem}\label{lemma1}
Let $\{u_k;k=1,2,...,m\}\subset \mathbb{R}^{n}$ for $n,m \ge 1$ and let $S$ be an $n\times n$ real symmetric positive-definite matrix. Define $U = \begin{bmatrix} u_1 & u_2 & \cdots & u_m  \end{bmatrix}$ and $A = I_m +U^T S^{-1} U$, where $I_m$ is the $m\times m$ identity matrix. Then for any $1\le i,j\le m$,
$$
u_i^T (S+UU^T)^{-1}u_j= \delta_{ij} + (-1)^{i+j+1} \frac{ M_{ji}}{\det(A)},
$$
where $\delta_{ij}$ is the Kronecker delta function, $M_{ij}$ is the $(i, j)$-minor of $A$ and we have adopted the standard notation $\det(A)$ for the determinant of $A$. In particular, when $m=1$,
$$
u_1^T (S+u_1u_1^T)^{-1}u_1=  1-\frac{1}{A}.
$$
\end{lem}

The following theorem gives a closed-form expression for the SK(1) capacity $C_{\mathrm{SK1}}(P)$ over stationary AR$(p)$ Gaussian channels, which provides a computable lower bound on the feedback capacity $C_\mathrm{FB}(P)$.
\begin{thm}\label{ISK1-thm}
Suppose the noise $\{ Z_i \}$ is a real stationary $AR(p)$ Gaussian process with power spectral density
\begin{equation*}
S_Z(e^{i\theta}) = \vert H_Z(e^{i\theta}) \vert^2 = \frac{1}{\vert L_Z(e^{i\theta}) \vert^2} = \frac{1}{\vert\prod_{k=1}^{p}(1+\beta_ke^{i\theta})\vert^2},
\end{equation*}
where $L_Z$ has real coefficients and $|\beta_k|<1$ for every $k$. Then, for every $P>0$,
\begin{equation}\label{I-SK1-thm}
C_\mathrm{SK1}(P)= \max_{\gamma} \log  |\gamma|,
\end{equation}
where the maximum is taken over all $\gamma\in\mathbb{R}$ with $|\gamma|>1$ satisfying the power constraint 
\begin{equation}\label{power-eqn-sk1-thm}
   (\gamma^2-1)H^2_Z(\gamma^{-1}) = P.
\end{equation}
\end{thm}
\begin{remark}\label{remark1:label}
For the special cases $p=0$ and $p=1$, corresponding to the AWGN channel and AR(1) Gaussian channel, respectively, the feedback capacity $C_{\mathrm{FB}}(P)$ under power constraint $P$ is already known from the existing literature (see \cite{kim2009feedback,yang2007feedback} and \cite[Theorem 5.2]{su2024feedback}). It follows directly from these results that $C_\mathrm{SK1}(P)=C_{\mathrm{FB}}(P)$ for both cases. Therefore, the SK(1) coding scheme is optimal for the AWGN and AR(1) Gaussian channels.
\end{remark}

\begin{proof}
The proof is given in Appendix~\ref{sec:proof_of_theorem_ref_isk1_thm} and uses the same observation-transform argument as the proof of Theorem~\ref{main-theorem-sk2}.
\end{proof}

\subsection{Butman's Equal-Energy Coding Scheme}\label{subsec:relation_to_butman}
In this subsection, we revisit Butman's equal-energy linear-feedback coding scheme in the notation of this paper \cite{butman1976linear}. We first recall the AR(1) case, where Butman's construction gives the known feedback capacity. We then identify sign and branch issues in the AR($p$) extension, and then adopt a sign-consistent, branch-complete formulation based on a real constant signed ratio. This formulation coincides with SK(1). We use it to state a corrected form of Butman's conjecture, whose failure for certain AR(2) Gaussian channels is established in Section~\ref{sec:sk2-coding-scheme}.

Butman extended the SK coding scheme from the AWGN channel to the ACGN channel \eqref{Channel-ACGN} (see \cite{butman1969,butman1976linear}). More precisely, the linear-feedback signaling rule can be formulated as
$$
    X_i=g_i\bigl(U-\mathbb E[U\mid Y_1^{i-1}]\bigr),
$$
where $U$ is a Gaussian message variable independent of the noise, and \(\{g_i\}\subset\mathbb R\) is chosen to satisfy average power constraint $P$. For the stationary AR(1) Gaussian channel with coefficient \(\beta_1\), Butman imposed an equal-energy allocation and a recursive sign selection. Define $r_i=|g_i/g_{i-1}|$. For $\beta_1\neq 0$, choose the sign of \(g_i\) such that
\begin{equation}
    \operatorname{sign}(g_i)=\operatorname{sign}(\beta_1 g_{i-1}).
\end{equation}
Then, Butman obtained
\[
    r_{i+1}^2 = 1+P\left(1+\frac{|\beta_1|}{r_i}\right)^2 .
\]
Under the equal-energy allocation \(r_i=x>1\), this gives
\begin{equation}\label{butman-original-rate:label}
    R^{(1)}_{ee}(P)=\log x,
\end{equation}
where $x$ is the unique positive root of the following polynomial equation
\[
    P=\frac{x^2-1}{(1+|\beta_1|x^{-1})^2}.
\]
This expression agrees with the known feedback capacity of the AR(1) Gaussian channel \cite{kim2009feedback,yang2007feedback,sabag2023}. When $\beta_1=0$, the channel is the AWGN channel and the nonzero gain signs may be chosen arbitrarily.

Butman further proposed an extension of this construction to the stationary AR($p$) Gaussian channel \eqref{channel-arp}. In the notation of this paper, we write
\[
    L_Z(z)=\prod_{k=1}^{p}(1+\beta_k z)
          =1+\sum_{k=1}^{p}b_k z^k
\]
for some $\{b_k;k=1,2,\cdots,p\}\subset \mathbb{R}$, and define
\[
    B(x)=\sum_{k=1}^{p}b_kx^{-k}.
\]
After the recursive sign selection and the equal-energy allocation, Butman's derivation leads to the following absolute-value equation
\begin{equation}\label{butman-abs-eqn}
    x^2= 1+ P\left(1+|B(x)|\right)^2 .
\end{equation}
Thus, if the equation \eqref{butman-abs-eqn} is valid, the corresponding achievable rate is
$$
    R^\mathrm{abs}_{ee}(P)=\frac12\log x^2=\log|x|,
$$
for any real root $x$ satisfying \eqref{butman-abs-eqn} and $|x|>1$.

Some issue occurs in the step from \eqref{butman-abs-eqn} to Butman's final polynomial expression (\cite[Eq. (49)]{butman1976linear}), which in the present notation corresponds to
\begin{equation}\label{butman-final-eqn-1}
    x^2= 1+ P\left(1-B(x)\right)^2.
\end{equation}
Butman's polynomial construction \cite[Eq. (49) and the discussion following it]{butman1976linear} selects the greatest real root $x_B$ of \eqref{butman-final-eqn-1} and assigns the rate
\begin{equation}\label{Butman-ARp-mutual-info}
    R_{ee}^{(p)}(P) = \log|x_B|.
\end{equation}
However, selecting the greatest real root does not, in general, maximize the modulus in \eqref{Butman-ARp-mutual-info}. The corrected benchmark below therefore retains both real branches. The step from \eqref{butman-abs-eqn} to \eqref{butman-final-eqn-1} replaces $|B(x)|$ by $-B(x)$, which is valid precisely when $B(x)\le 0$.

Butman also noted that ``A little thought will reveal that the above choice of sign for \(g_n\), may not be the best. This problem deserves more investigation. However, if all the \(\alpha\) have a negative sign, the above choice is the best.'' Note that $\alpha_k$ in \cite{butman1976linear} corresponds to $-b_k$ in this paper. Hence, Butman's sufficient condition that all \(\alpha_k\) are negative corresponds to
$$
    b_k>0,\qquad k=1,\ldots,p .
$$
Under this coefficient condition, $B(x)>0$ on the positive branch \(x>0\), so $|B(x)|=B(x)$. Consequently, on this positive branch $x>0$, the sign-consistent equation following from \eqref{butman-abs-eqn} is not \eqref{butman-final-eqn-1}, but rather
\begin{equation}\label{butman-final-eqn-2}
    x^2=1+P\left(1+B(x)\right)^2 .
\end{equation}
Thus, on this positive branch $x>0$, the corresponding achievable rate is
$$
    R^{(+)}_{\mathrm{Butman}}(P)=\max \log x,
$$
where the maximum is taken over all real roots $x$ of \eqref{butman-final-eqn-2} satisfying $x>1$. Therefore, the transition from \eqref{butman-abs-eqn} to \eqref{butman-final-eqn-1} is not valid in general; the correct sign depends on the value of $B(x)$ at the selected root $x$.

There is a further compatibility issue behind \eqref{butman-abs-eqn}. In Butman's derivation for the AR($p$) case, the ratios $x_i=g_i/g_{i-1}$ are signed quantities. Thus, the equal-energy allocation $x_i=x$, \(|x|>1\), restricts the sign selection of \(\{g_i\}\). Indeed, let $s_i=\operatorname{sign}(g_i) \in \{1,-1\}$. If the additional constant signed-ratio restriction $x_i=x$ is imposed, then
\begin{equation}\label{Butman-ee-cond}
    s_i=\operatorname{sign}(x)s_{i-1}.
\end{equation}
Thus, the sign sequence $\{s_i\}$ is constant when $x>0$ and alternating when $x<0$. Furthermore, with $r=|x|$, the squared term in Butman's recursion (see \cite[Eq. (39)]{butman1976linear}) can be written as 
\begin{equation}\label{Butman-recursive-term}
    \left(1+\sum_{k=1}^{p}b_k\frac{g_{i-k}}{g_i}\right)^2 = \left(1+s_i\sum_{k=1}^{p}b_k s_{i-k}r^{-k}\right)^2.
\end{equation}
Define
\[
    S_i(r)=\sum_{k=1}^{p} b_k s_{i-k} r^{-k}.
\]
For fixed $\{s_{i-1},\cdots,s_{i-p}\}$, the two possible choices $s_i=1$ and $s_i=-1$ give $(1+S_i(r))^2$ and $(1-S_i(r))^2$, respectively. Therefore, to maximize \eqref{Butman-recursive-term}, when \(S_i(r)\neq0\), Butman's recursive sign choice is
\begin{equation}\label{Butman-sign-cond}
    s_i=\operatorname{sign}(S_i(r)).
\end{equation}
When \(S_i(r)=0\), either sign may be chosen. The squared term \eqref{Butman-recursive-term} becomes
\begin{equation}\label{Butman-one-step-factor}
    \left(1+|S_i(r)|\right)^2.
\end{equation}

On the other hand, it follows from \eqref{Butman-ee-cond} that
\[
    s_{i-k}=s_i\operatorname{sign}(x)^k,
\]
and hence
\[
    S_i(r) = s_i\sum_{k=1}^{p} b_k x^{-k} = s_i B(x).
\]
When $B(x)\neq 0$, the recursive sign-selection rule \eqref{Butman-sign-cond} is compatible with \eqref{Butman-ee-cond} if and only if $\operatorname{sign}(B(x))=1$. When $B(x)=0$, either sign may be chosen. Thus, the necessary and sufficient compatibility condition is
\begin{equation}\label{Butman-branch-2}
    B(x)\ge 0.
\end{equation}
Hence, the equation \eqref{butman-abs-eqn} is not valid without this compatibility condition \eqref{Butman-branch-2}. 

For $p=1$, this compatibility condition \eqref{Butman-branch-2} is automatic. Indeed,
$$
    B(x)=b_1x^{-1}.
$$
Given \(r>1\), choose $x=\operatorname{sign}(b_1)r$ when \(b_1\neq0\). Then
\[
    B(x)=\frac{|b_1|}{r}>0.
\]
Moreover,
\[
    S_i(r)=b_1s_{i-1}r^{-1},
\]
such that
\[
    \operatorname{sign}(S_i(r)) = \operatorname{sign}(b_1)s_{i-1} = \operatorname{sign}(x)s_{i-1} = s_i.
\]
Thus, \eqref{Butman-sign-cond} and \eqref{Butman-ee-cond} are consistent in the AR(1) case, and \eqref{Butman-one-step-factor} becomes
\[
    \left(1+|S_i(r)|\right)^2 = \left(1+\frac{|b_1|}{r}\right)^2,
\]
which is Butman's AR(1) recursion (\cite[Eq. (4)]{butman1976linear}). 

For $p>1$, however, a real root $x$ of the absolute-value equation \eqref{butman-abs-eqn} need not satisfy the compatibility condition \eqref{Butman-branch-2}. Example~\ref{Butman-example-1:label} in Appendix~\ref{sec:example_of_ar_2_gaussian_channel} gives an AR(2) channel for which the absolute-value equation \eqref{butman-abs-eqn} has real roots but every such root has $B(x)<0$, so none is compatible. Thus, \eqref{butman-abs-eqn} is not, by itself, a generally valid achievable-rate characterization for AR($p$) Gaussian channels; an additional sign-compatibility condition is required. Separately, Example~\ref{Butman-example-2:label} in Appendix~\ref{sec:example_of_ar_2_gaussian_channel} gives an AR(2) channel for which $B(x)>0$ for every real $x\neq 0$, so the fixed replacement $|B(x)|=-B(x)$ leading to the polynomial equation \eqref{butman-final-eqn-1} is invalid. Examples~\ref{Butman-example-1:label} and \ref{Butman-example-2:label} show that the AR($p$) fixed-point derivation in \cite{butman1976linear} does not, as written, justify the asserted formula in general. The first example concerns compatibility with a real constant signed ratio, while the second concerns the fixed-sign removal of an absolute value. These issues motivate the corrected formulation below.

We therefore return to the exact recursion \cite[Eq. (39)]{butman1976linear} and retain the constant signed-ratio specialization adopted after \cite[Eq. (42)]{butman1976linear}. Let $x_i=g_i/g_{i-1}=x\in\mathbb{R}$, $|x|>1$. The gain signs are determined by this ratio, and no separate greedy sign-selection rule is imposed. Butman's equation \cite[Eq. (39)]{butman1976linear} becomes
\begin{align}
    x^2  = 1+P\left(1+\sum_{k=1}^{p}b_k x^{-k}\right)^2= 1+P L_Z^2(x^{-1}). \label{butman-corrected-signed-eqn}
\end{align}
Equivalently,
\begin{equation}\label{butman-corrected-power}
    P
    =
    \frac{x^2-1}{L_Z^2(x^{-1})}
    =
    \frac{x^2-1}
    {\left(1+\sum_{k=1}^{p}b_kx^{-k}\right)^2}.
\end{equation}
In particular, when \(x>1\), \eqref{butman-corrected-signed-eqn} reduces to the positive branch \eqref{butman-final-eqn-2} and the corresponding achievable rate is exactly $R^{(+)}_{\mathrm{Butman}}(P)$, whereas the case \(x<-1\) is obtained by evaluating \(L_Z(x^{-1})\) at a negative argument.

Thus, the resulting corrected constant signed-ratio benchmark is given by
\begin{equation}\label{butman-branch-complete-rate}
R_{\mathrm{Butman}}(P) = \max_\gamma\log |\gamma|,
\end{equation}
where the maximum is taken over all $\gamma\in\mathbb{R}$ satisfying $|\gamma|>1$ and 
\begin{equation}\label{butman-branch-complete-power}
   (\gamma^2-1)H_Z^2(\gamma^{-1})=P,
\end{equation}
where \(H_Z(z)=1/L_Z(z)\). Combining \eqref{butman-branch-complete-rate}--\eqref{butman-branch-complete-power} with Theorem~\ref{ISK1-thm}, we obtain
\begin{equation}\label{butman-sk1-identity}
    R_{\mathrm{Butman}}(P)=C_{\mathrm{SK1}}(P).
\end{equation}
Thus, SK(1) provides a sign-consistent, branch-complete formulation of the real constant signed-ratio specialization of Butman's scalar linear-feedback construction, allowing both $\gamma>1$ and $\gamma<-1$.

Butman \cite{butman1976linear} conjectured that the equal-energy lower bound approaches feedback capacity asymptotically. In view of the sign and branch issues discussed above, we interpret Butman's conjecture through the corrected constant signed-ratio formulation \eqref{butman-branch-complete-rate}--\eqref{butman-sk1-identity} and state the resulting optimality claim below.

\begin{conj}[A corrected form of Butman's conjecture]\label{Butman-conj:label}
For the stationary AR($p$) Gaussian channels \eqref{channel-arp}--\eqref{sdf-arp} and every $P>0$,
\[
    C_{\mathrm{FB}}(P)=C_{\mathrm{SK1}}(P).
\]
\end{conj}
For $p=1$, this corrected conjecture holds by the known feedback-capacity formula for ARMA(1,1) Gaussian channels \cite{kim2009feedback,yang2007feedback,sabag2023}. For \(p>1\), however, it is false in general. Proposition~\ref{prop:interleaved-AR1} gives an explicit interleaved family of stationary AR(2) Gaussian channels and establishes
$$
C_{\mathrm{FB}}(P)=C_{\mathrm{SK2}}(P)>C_{\mathrm{SK1}}(P),
$$
thereby disproving Conjecture~\ref{Butman-conj:label}. The numerical examples in Section~\ref{subsec:numerical_analysis} further illustrate strict improvements of SK(2) over SK(1) beyond this interleaved family.

In the following section, we introduce the SK(2) coding scheme, in which the underlying message process follows a second-order deterministic recursion. This provides a strict generalization of the SK(1) coding scheme.

\section{The SK(2) Coding Scheme}\label{sec:sk2-coding-scheme}
Building on the framework of the SK(1) coding scheme, we now present its natural extension: the SK(2) coding scheme. The message process now follows a second-order deterministic recursion. Formally, let $U_1$ and $U_2$ be two independent standard normal random variables, independent of the noise process $\{Z_i\}$. The transmitter initially sends 
$$
X_1 = U_1,\qquad X_2 = U_2
$$
and subsequently sends 
\begin{equation}\label{SK2-coding-input}
X_i = V_i - \E{[V_i\mid Y_1^{i-1}]}, \qquad i=3,4,\dots
\end{equation}
where $\{V_n\}$ is the message process satisfying 
\begin{equation}\label{SK2-coding-message}
    V_1=U_1,\qquad V_2=U_2,\qquad V_{i+1} = aV_i + bV_{i-1},\quad i\ge2,
\end{equation}
where $a,b\in\mathbb{R}$. Analogous to the definition of the SK(1) capacity in \eqref{feedback-capacity-SK1}, we define the {\em SK(2) information rate capacity}, or simply, {\em SK(2) capacity} as
\begin{equation*}
C_{\mathrm{SK2}}(P) = \sup_{(V,X)} \bar{I}(V;Y),
\end{equation*}
where the supremum is taken over all pairs $(V,X)$ satisfying (\ref{SK2-coding-input}), (\ref{SK2-coding-message}) and the constraint \eqref{power-constraint-sk1}. Thus, $C_{\mathrm{SK2}}(P)$ is the maximal achievable information rate
within the SK(2) coding class. In particular,
$$
    C_{\mathrm{SK2}}(P)\le C_{\mathrm{FB}}(P).
$$

Moreover, when $b=0$ and $|a|>1$, the recursion \eqref{SK2-coding-message} reduces to the first-order recursion $V_{n+1}=aV_n$ after initialization. Appendix~\ref{sec-app:reduction_lemma} shows that the mutual information rate and input power coincide with those of the SK(1) scheme with $\gamma=a$. Thus, the SK(1) coding scheme is essentially included in the SK(2) coding class as a special case, and consequently
\[
    C_{\mathrm{SK1}}(P)\le C_{\mathrm{SK2}}(P)\le C_{\mathrm{FB}}(P).
\]

The full SK(2) class allows arbitrary real coefficients $a$ and $b$ in \eqref{SK2-coding-message}. Let $\gamma_1$ and $\gamma_2$ be the roots of the corresponding characteristic equation (in $\lambda$) of \eqref{SK2-coding-message}:
\begin{equation}\label{sk2-char-eqn}
   \lambda^2 - a\lambda - b =0.
\end{equation}
A characteristic root is called {\em unstable} if its modulus is strictly greater than one. Throughout, roots are counted with algebraic multiplicity. Thus, the full class includes recursions with zero, one, or two unstable roots. We first analyze the case of two unstable roots.

Define the set of admissible characteristic pairs by
\begin{equation}\label{def-admissible-pair}
 \mathcal{A}_2 \triangleq \{(\gamma_1, \gamma_2) \in \mathbb{C}^2 : |\gamma_1| \wedge |\gamma_2| > 1, \ \gamma_1 + \gamma_2 \in \mathbb{R}, \ \gamma_1\gamma_2 \in \mathbb{R}\},
\end{equation}
where $\phi\wedge \psi\triangleq \min(\phi,\psi)$. A characteristic pair $(\gamma_1,\gamma_2)$ is called {\em admissible} if $(\gamma_1,\gamma_2)\in\mathcal A_2$. An SK(2) recursion pair $(V,X)$ satisfying \eqref{SK2-coding-input}--\eqref{SK2-coding-message} is called {\em genuinely second-order} if its characteristic pair is admissible. Thus, a genuine second-order recursion has two unstable characteristic roots. Each admissible pair $(\gamma_1,\gamma_2)$ determines the real coefficients $a=\gamma_1+\gamma_2$ and $b=-\gamma_1\gamma_2$ in \eqref{SK2-coding-message}. Admissibility implies $|b|=|\gamma_1\gamma_2|>1$, whereas neither $b\neq 0$ nor $|b|>1$ alone ensures admissibility. An admissible pair $(\gamma_1,\gamma_2)$ is {\em feasible} under power constraint $P$ if $P_2(\gamma_1,\gamma_2)\le P$, where $P_2(\gamma_1,\gamma_2)$ is the asymptotic input power specified in Theorem~\ref{main-theorem-sk2}.

Then, $\{V_n\}$ admits the finite-rank representation
\begin{equation*}\label{Vn-genaral-sol}
   V_n = a_n U_1 +b_n U_2,\quad n\ge1,
\end{equation*}
where the deterministic coefficients $(a_n,b_n)$ are given as follows, according to the nature of the roots.

\textbf{Case A:} $\gamma_1 \neq \gamma_2$ (distinct roots). Then 
\begin{equation}\label{Vn-case-a}
   a_n = -(c_1\gamma_2\gamma_1^n +c_2\gamma_1\gamma_2^n),\qquad b_n = c_1\gamma_1^n +c_2\gamma_2^n,
\end{equation}
where
$$
   c_1 = \frac{1}{\gamma_1(\gamma_1-\gamma_2)},\qquad c_2 = \frac{1}{\gamma_2(\gamma_2-\gamma_1)}.
$$

\textbf{Case B:} $\gamma_1 = \gamma_2 =: \gamma$ (repeated root). Then
\begin{equation}\label{Vn-case-b}
   a_n = (2-n)\gamma^{n-1},\qquad b_n = (n-1)\gamma^{n-2}.
\end{equation}

We define the {\em genuine SK(2) capacity} under power constraint $P$ by
\begin{equation}\label{def-gen-sk2-cap}
C^{\mathrm{g}}_{\mathrm{SK2}}(P)=\sup_{(V,X)}\bar{I}(V;Y),
\end{equation}
where the supremum is taken over all genuinely second-order pairs $(V,X)$ satisfying \eqref{SK2-coding-input}, \eqref{SK2-coding-message}, and \eqref{power-constraint-sk1}. In particular, $b\neq 0$ is necessary but not sufficient for genuine second order.

\subsection{Characterization of the SK(2) Capacity} 
\label{sub:sk_2_capacity}
The following theorem provides a closed-form characterization of the genuine SK(2) capacity $C^\mathrm{g}_\mathrm{SK2}(P)$. Its relation to the full SK(2) class is established in Lemma~\ref{lemma2:label} and Corollary~\ref{main-col-sk2:label} below. For later use, we define $R_2(\gamma_1,\gamma_2)\triangleq \log |\gamma_1\gamma_2|$ for $(\gamma_1,\gamma_2)\in\mathcal A_2$.

\begin{thm}\label{main-theorem-sk2}
Suppose that $\{ Z_i \}$ is a stationary AR$(p)$ Gaussian process with the power spectral density \eqref{sdf-arp}, where $|\beta_k|<1$ for $k=1,\ldots,p$ and $L_Z$ has real coefficients. For every $P>0$,
\begin{equation}\label{I-SK2-thm}
C^\mathrm{g}_{\mathrm{SK2}}(P) =\sup_{\gamma_1,\gamma_2} \log  |\gamma_1\gamma_2|,
\end{equation}
where the supremum is taken over all pairs satisfying
\begin{equation}\label{I-SK2-thm-constraints}
(\gamma_1,\gamma_2)\in \mathcal{A}_2,\qquad P_2(\gamma_1,\gamma_2)= P,
\end{equation}
and $\mathcal A_2$ is the set of admissible characteristic pairs specified by \eqref{def-admissible-pair} and $P_2(\gamma_1,\gamma_2)$ is given by
\begin{equation}\label{power-eqn-sk2-thm}
P_2(\gamma_1,\gamma_2)=
\begin{cases}
\displaystyle
 \frac{1}{\Delta}
 \left(
    \frac{L_{11}}{\gamma_2^2} +\frac{L_{22}}{\gamma_1^2} +\frac{2L_{12}}{\gamma_1\gamma_2} \right),
 & \gamma_1\ne\gamma_2,\\[16pt]
\begin{aligned}
 &\frac{(\gamma^2-1)^3[L_Z'(\gamma^{-1})]^2}
        {\gamma^4L_Z(\gamma^{-1})^4}+\frac{2(\gamma^2-1)^2L_Z'(\gamma^{-1})}
              {\gamma L_Z(\gamma^{-1})^3}
       +\frac{\gamma^4-1}{L_Z(\gamma^{-1})^2},
\end{aligned} & \gamma_1=\gamma_2=\gamma.
\end{cases}
\end{equation}
For distinct roots, the quantities in the first case are defined by
\begin{equation}\label{main-thm-notations}
\begin{aligned}
\Delta &= (\gamma_1-\gamma_2)^2(L_{11}L_{22}-L_{12}^2),\\
L_{ij} &=\frac{c_{i}c_{j}L_Z(\gamma_i^{-1})L_Z(\gamma_j^{-1})}{\gamma_i\gamma_j-1}, \qquad \mbox{$i,j=1,2,$}\\
c_1 &= \frac{1}{\gamma_1(\gamma_1-\gamma_2)}, \quad c_2 = \frac{1}{\gamma_2(\gamma_2-\gamma_1)}.
\end{aligned}
\end{equation}
Here $L_Z^\prime$ denotes the derivative of $L_Z$.
\end{thm}

\begin{remark}\label{remark-continuous-extension:label}
In the repeated-root case $\gamma_1=\gamma_2=\gamma$ of \eqref{power-eqn-sk2-thm}, admissibility implies $\gamma\in\mathbb R$ and $|\gamma|>1$, and moreover, $P_2(\gamma,\gamma)$ is the joint continuous extension of the distinct-root expression $\gamma_1\neq \gamma_2$, including approaches through complex conjugate pairs.
\end{remark}

\begin{remark}\label{remark-exact-Power:label}
The constraint $P_2(\gamma_1,\gamma_2)=P$ in \eqref{I-SK2-thm-constraints} may equivalently be replaced by $P_2(\gamma_1,\gamma_2)\le P$. Indeed, for a fixed admissible pair $(\gamma_1,\gamma_2)$ with $P_2(\gamma_1,\gamma_2)<P$, it holds that the pair $(c\gamma_1,c\gamma_2)$ remains admissible for any real $c \ge 1$. Clearly, as a function of $c$, $P_2(c\gamma_1,c\gamma_2)$ is continuous. By \eqref{power-eqn-sk2-thm} and $L_Z(0)=1$, we have
$$
\lim_{c\to\infty}\frac{P_2(c\gamma_1,c\gamma_2)}{c^4}=|\gamma_1\gamma_2|^2>0.
$$
By the intermediate value theorem, there exists a real $c^\ast\ge 1$ such that $(c^\ast\gamma_1,c^\ast\gamma_2)$ is admissible and $P_2(c^\ast\gamma_1,c^\ast\gamma_2)=P$. Moreover,
$$
R_2(c^\ast\gamma_1,c^\ast\gamma_2)=R_2(\gamma_1,\gamma_2)+2\log c^\ast\ge R_2(\gamma_1,\gamma_2).
$$
Thus, every pair with strict power slack is dominated by an equality-power pair, and the two constraints give the same supremum.
\end{remark}

\begin{proof}
Consider the SK(2) coding scheme formulated by \eqref{SK2-coding-input} and \eqref{SK2-coding-message} over the stationary AR$(p)$ Gaussian channel $Y_i = X_i +Z_i$ with the noise spectral density function $S_Z(e^{i\theta})$ given by (\ref{sdf-arp}).

We first address Case A. Recall that 
\begin{equation*}
\begin{aligned}
Y_n &= X_n + Z_n= V_n -\E[V_n\mid Y_1^{n-1}] +Z_n
\end{aligned}
\end{equation*}
and define
\begin{equation*}
Y_n^\ast = V_n + Z_n.
\end{equation*}
Clearly, $Y_n^\ast = Y_n +\E[V_n\mid Y_1^{n-1}]$ and therefore, for any $i\ge 1$, $\{ Y_1^i \}$ and $\{Y_1^{\ast,i}\}$ are uniquely determined by each other. Furthermore, define
\begin{equation}\label{Y-tlta}
\widetilde{Y}_n =\left\{ 
\begin{aligned}
& Y_n^\ast, ~~~~~~~~~~ n\le p;\\[2pt]
&L(B)Y^\ast_n, ~~~n\ge p+1,
\end{aligned}
\right.
\end{equation}
where $L$ stands for $L_Z$, $B$ denotes the backward-shift operator. Now by (\ref{Vn-case-a}), we can express 
\begin{equation}\label{Y-tlta-2}
\begin{aligned}
\widetilde{Y}_n & = L(B)(V_n +Z_n)\\
& =  L(B)(a_n U_1 +b_nU_2) +L(B)Z_n\\
&= d_{1,n}U_1 +d_{2,n}U_2 +W_n
\end{aligned}
\end{equation}
for $n\ge p+1$, where
\begin{equation}\label{d1d2}
\left\{ 
\begin{aligned}
d_{1,n} &=-c_1\gamma_2\gamma_1^nL(\gamma_1^{-1})- c_2\gamma_1\gamma_2^n L(\gamma_2^{-1});\\[2pt]
d_{2,n} &= c_1\gamma_1^nL(\gamma_1^{-1}) + c_2\gamma_2^n L(\gamma_2^{-1}).
\end{aligned}
\right.
\end{equation}
Thus, by \eqref{Y-tlta}, \eqref{Y-tlta-2} and \eqref{d1d2}, we have 
\begin{equation}\label{Y-tlta-re}
\widetilde{Y}_n =\left\{ 
\begin{aligned}
& a_n U_1 +b_n U_2 + Z_n, ~~~~~~~~~~~~ n\le p;\\[2pt]
&d_{1,n}U_1 + d_{2,n}U_2 + W_n,~~~~ ~~~n\ge p+1.
\end{aligned}
\right.
\end{equation}

For $n\ge \max \{p+1,2\}$, we have the following chain of equalities:
\begin{equation}\label{mutual-info-sk2}
\begin{aligned}
\frac{1}{n}I(V_1^n;Y_1^n) &\overset{(a)}{=}\frac{1}{n}I(U_1,U_2;Y_1^n)\\
&\overset{(b)}{=}\frac{1}{n}I(U_1,U_2;\widetilde{Y}_1^n)      \\
&= \frac{1}{n}I(U_1;\widetilde{Y}_1^n) +\frac{1}{n}I(U_2;\widetilde{Y}_1^n\mid U_1)\\
&\overset{(c)}{=}-\frac{1}{2n}\log\E[\vert U_1-\E[U_1\mid \widetilde{Y}_{1}^n] \vert^2] -\frac{1}{2n}\log\E[\vert U_2-\E[U_2\mid U_1, \widetilde{Y}_{1}^n] \vert^2],
\end{aligned}
\end{equation}
where $(a)$ follows from \eqref{SK2-coding-message}, $(b)$ follows from the fact that $Y_1^n$ and $\widetilde{Y}_1^n$ are uniquely determined by each other and $(c)$ follows from the relation between mutual information and MMSE for Gaussian random variables, together with the entropy formula for a multivariate normal distribution (see, e.g., \cite[Theorem 8.6.6 \& its corollary]{cover1991elements}).

Next, for $n\ge p+1$, we also have
\begin{equation}\label{power-eqn-sk2}
   \begin{aligned}
      \E[|X_n|^2] = \E[\vert V_n-\E[V_n\mid Y_1^{n-1}]\vert ^2]\overset{(d)}{=}\E[\vert V_n - \E[ V_n\mid \widetilde{Y}_1^{n-1}]\vert^2],
   \end{aligned}
\end{equation}
where $(d)$ follows from the fact that $Y_1^{n-1}$ and $\widetilde{Y}_1^{n-1}$ are uniquely determined by each other.

Let 
\[
A = \begin{bmatrix} a_1 & a_2 & \cdots & a_p \end{bmatrix}^T,\qquad
B = \begin{bmatrix} b_1 & b_2 & \cdots & b_p \end{bmatrix}^T,
\]
and
\[
D_{1,n} = \begin{bmatrix} d_{1,p+1} & d_{1,p+2} & \cdots & d_{1,n} \end{bmatrix}^T,\qquad
D_{2,n} = \begin{bmatrix} d_{2,p+1} & d_{2,p+2} & \cdots & d_{2,n} \end{bmatrix}^T.
\]
Since $U_1$ and $\{\widetilde{Y}_1^n\}$ are jointly Gaussian, the conditional mean squared error is given by 
\begin{equation}\label{MMSE-U-1-a}
\begin{aligned}
   \E[\vert U_1-\E[U_1\mid \widetilde{Y}_1^n] \vert^2] &= \mathrm{Var}(U_1) - \mathrm{Cov}(U_1,\widetilde{Y}_1^n)(\mathrm{Var}(\widetilde{Y}_1^n))^{-1}\mathrm{Cov}(\widetilde{Y}_1^n,U_1)\\
   &= 1 - \begin{bmatrix}A^T & D_{1,n}^T \end{bmatrix}\Sigma_{\widetilde{Y}_1^n}^{-1}\begin{bmatrix}A\\[2pt] D_{1,n}\end{bmatrix},
\end{aligned}
\end{equation}
where the second equality uses $\mathrm{Var}(U_1)=1$ and the fact that
\[
   \mathrm{Cov}(U_1,\widetilde{Y}_1^n)= \begin{bmatrix}A^T & D_{1,n}^T \end{bmatrix}.
\]
Here $\Sigma_{\widetilde{Y}_1^n} \triangleq\mathrm{Var}(\widetilde{Y}_1^n)$ is given explicitly by
\begin{equation}\label{def-Sigma-Y-tilta}
   \Sigma_{\widetilde{Y}_1^n} =\begin{bmatrix}
         AA^T+BB^T+\Sigma_Z & A D_{1,n}^T+BD_{2,n}^T\\[2pt]
         D_{1,n}A^T +D_{2,n} B^T & I_{n-p}+D_{1,n}D_{1,n}^T +D_{2,n}D_{2,n}^T
      \end{bmatrix},
\end{equation}
where $\Sigma_Z\triangleq\mathrm{Var}(Z_1^p)$, and $I_{n-p}$ denotes the $(n-p)\times(n-p)$ identity matrix. 

Moreover, let 
\[
   \widetilde{D}_{1,n}=\begin{bmatrix}A \\[2pt]D_{1,n}\end{bmatrix}, \qquad\widetilde{D}_{2,n}=\begin{bmatrix}B\\[2pt]D_{2,n}\end{bmatrix},\qquad\widetilde{\Sigma}_{Z,n}=\begin{bmatrix}\Sigma_Z&0\\[2pt]0 &I_{n-p}\end{bmatrix}.
\]
Then, the equation \eqref{MMSE-U-1-a} can be rewritten in the following compact form: 
\begin{equation}\label{MMSE-U-1}
\begin{aligned}
   \E[\vert U_1-\E[U_1\mid \widetilde{Y}_1^n] \vert^2] &=1 - \widetilde{D}_{1,n}^T\left(\widetilde{\Sigma}_{Z,n} +\widetilde{D}_{1,n}\widetilde{D}_{1,n}^T +\widetilde{D}_{2,n}\widetilde{D}_{2,n}^T\right)^{-1}\widetilde{D}_{1,n}.
\end{aligned}
\end{equation}
Note that by \cite[Proposition 5.1.1]{brockwell2009time}, $\Sigma_Z$ is non-singular and hence positive definite. Then, applying Lemma \ref{lemma1} to \eqref{MMSE-U-1} yields 
\begin{equation}\label{MMSE-U-2}
\begin{aligned}
   \E[\vert U_1-\E[U_1\mid \widetilde{Y}_1^n]\vert^2]&=\frac{1+\widetilde{D}_{2,n}^T\widetilde{\Sigma}_{Z,n}^{-1}\widetilde{D}_{2,n}}{\det\begin{pmatrix}
      1+\widetilde{D}^T_{1,n}\widetilde{\Sigma}_{Z,n}^{-1}\widetilde{D}_{1,n} & \widetilde{D}^T_{1,n}\widetilde{\Sigma}_{Z,n}^{-1}\widetilde{D}_{2,n}\\[2pt]
      \widetilde{D}^T_{2,n}\widetilde{\Sigma}_{Z,n}^{-1}\widetilde{D}_{1,n} & 1+ \widetilde{D}^T_{2,n} \widetilde{\Sigma}_{Z,n}^{-1}\widetilde{D}_{2,n}
   \end{pmatrix}}\\
   &=\frac{1+ B^T\Sigma_Z^{-1}B +D_{2,n}^TD_{2,n}}{\det\begin{pmatrix}1+A^T\Sigma_Z^{-1}A +D^T_{1,n}D_{1,n} & A^T\Sigma_Z^{-1}B+D^T_{1,n}D_{2,n}\\[2pt]
      B^T\Sigma_Z^{-1}A+D^T_{2,n}D_{1,n} &1+B^T\Sigma_Z^{-1}B+D^T_{2,n}D_{2,n}
   \end{pmatrix}}.
\end{aligned}
\end{equation}
Similarly, for $U_2$ we obtain
\begin{equation}\label{MMSE-V-2}
\begin{aligned}
\E[\vert U_2-\E[U_2\mid U_1, \widetilde{Y}_1^n]\vert^2] &= 1 - \begin{bmatrix}0 & \widetilde{D}^T_{2,n}\end{bmatrix} \begin{bmatrix}1 & \widetilde{D}_{1,n}^T\\[2pt] \widetilde{D}_{1,n} & \widetilde{\Sigma}_{Z,n}+\widetilde{D}_{1,n}\widetilde{D}_{1,n}^T+\widetilde{D}_{2,n}\widetilde{D}_{2,n}^T\end{bmatrix}^{-1} \begin{bmatrix}
   0\\ \widetilde{D}_{2,n}
\end{bmatrix}        \\
&= 1 -   \widetilde{D}_{2,n}^T(\widetilde{\Sigma}_{Z,n}+\widetilde{D}_{2,n}\widetilde{D}_{2,n}^T)^{-1}\widetilde{D}_{2,n}         \\
&=\frac{1}{1+\widetilde{D}_{2,n}^T\widetilde{\Sigma}_{Z,n}^{-1}\widetilde{D}_{2,n}}\\
&=\frac{1}{1+B^T\Sigma_Z^{-1}B+D^T_{2,n}D_{2,n}}.
\end{aligned}
\end{equation}
Consequently, it follows from \eqref{MMSE-U-2} and \eqref{MMSE-V-2} that \eqref{mutual-info-sk2} can be simplified as
\begin{equation}\label{pf-sk2-mutual-info}
   \frac{1}{n}I(V_1^n;Y_1^n)= \frac{1}{2n}\log \det(\Delta_{2,n}),
\end{equation}
where $\Delta_{2,n}=I_2 +\widetilde{D}_n^T\widetilde{\Sigma}_{Z,n}^{-1}\widetilde{D}_n$ and $\widetilde{D}_n =\begin{bmatrix}\widetilde{D}_{1,n} & \widetilde{D}_{2,n}\end{bmatrix}$.

On the other hand, since $V_n$ and $\{\widetilde{Y}_1^{n-1}\}$ are also jointly Gaussian, it follows from \cite[Theorem 4.6]{gray2004introduction} that
\begin{equation}\label{MMSE-Vn}
\begin{aligned}
\E[| V_n-\E[V_n\mid \widetilde{Y}_1^{n-1}]|^2]&= \mathrm{Var}(V_n) - \mathrm{Cov}(V_n,\widetilde{Y}_1^{n-1})\left(\mathrm{Var}(\widetilde{Y}_1^{n-1})\right)^{-1}\mathrm{Cov}(\widetilde{Y}_1^{n-1},V_n) \\
&= a_n^2 +b_n^2 -\begin{bmatrix}a_n&b_n\end{bmatrix}\widetilde{D}^T_{n-1}\Sigma_{\widetilde{Y}_1^{n-1}}^{-1}\widetilde{D}_{n-1}\begin{bmatrix}a_n\\[2pt]b_n\end{bmatrix},
\end{aligned}
\end{equation}
where the second equality uses $\mathrm{Var}(V_n)=a_n^2+b_n^2$ and the fact that
\begin{equation*}
\begin{aligned}
   \mathrm{Cov}(V_n,\widetilde{Y}_1^{n-1}) &= \begin{bmatrix}a_n A^T+b_nB^T& a_nD_{1,n-1}^T+b_nD_{2,n-1}^T\end{bmatrix}\\
   &=\begin{bmatrix}a_n &b_n \end{bmatrix}\begin{bmatrix}\widetilde{D}^T_{1,n-1}\\[2pt]\widetilde{D}^T_{2,n-1}\end{bmatrix}\\
   &=\begin{bmatrix}a_n&b_n\end{bmatrix}\widetilde{D}^T_{n-1}.
\end{aligned}
\end{equation*}
Now, by \eqref{def-Sigma-Y-tilta} with $n$ replaced by $n-1$, the covariance matrix of $\widetilde{Y}_1^{n-1}$ can be expressed as
\begin{equation*}
\begin{aligned}
   \Sigma_{\widetilde{Y}_1^{n-1}}&=\widetilde{\Sigma}_{Z,n-1}+\widetilde{D}_{1,n-1}\widetilde{D}^T_{1,n-1}+\widetilde{D}_{2,n-1}\widetilde{D}^T_{2,n-1}\\
&=\widetilde{\Sigma}_{Z,n-1}+\widetilde{D}_{n-1}\widetilde{D}^T_{n-1},
\end{aligned}
\end{equation*}
which, together with \eqref{MMSE-Vn}, yields
\begin{equation}\label{MMSE-Vn-final}
   \begin{aligned}
      \E[\vert& V_n-\E[V_n\mid \widetilde{Y}_1^{n-1}]\vert^2]=a_n^2+b_n^2-\begin{bmatrix}a_n&b_n\end{bmatrix}\widetilde{D}^T_{n-1}\left(\widetilde{\Sigma}_{Z,n-1}+\widetilde{D}_{n-1}\widetilde{D}^T_{n-1}\right)^{-1}\widetilde{D}_{n-1}\begin{bmatrix}a_n\\[2pt]b_n\end{bmatrix}\\ 
&\overset{(e)}{=}a_n^2+b_n^2-\begin{bmatrix}a_n &b_n\end{bmatrix}\left(I_2 - (I_2 +\widetilde{D}^T_{n-1}\widetilde{\Sigma}_{Z,n-1}^{-1}\widetilde{D}_{n-1})^{-1}\right)\begin{bmatrix}a_n\\[2pt]b_n\end{bmatrix}\\
&=\begin{bmatrix}a_n&b_n\end{bmatrix}\left(I_2 +\widetilde{D}^T_{n-1}\widetilde{\Sigma}_{Z,n-1}^{-1}\widetilde{D}_{n-1}\right)^{-1}      \begin{bmatrix}a_n\\[2pt]b_n\end{bmatrix}\\
&=\begin{bmatrix}a_n&b_n\end{bmatrix}(\Delta_{2,n-1})^{-1} \begin{bmatrix}a_n\\[2pt]b_n\end{bmatrix},
\end{aligned}
\end{equation}
where $(e)$ follows from Lemma \ref{lemma1}. Now, note that
\begin{equation*}
\begin{aligned}
\Delta_{2,n-1}&=\begin{bmatrix}1&0\\[2pt]0&1\end{bmatrix}+ \begin{bmatrix}A^T&D_{1,n-1}^T\\[2pt]B^T&D_{2,n-1}^T \end{bmatrix}
\begin{bmatrix}\Sigma_Z^{-1}&0\\[2pt]0&I_{n-p-1}\end{bmatrix}\begin{bmatrix}A&B\\[2pt]D_{1,n-1}&D_{2,n-1}\end{bmatrix}\\
&=\begin{bmatrix}
1+A^T\Sigma_Z^{-1}A +D_{1,n-1}^TD_{1,n-1} & A^T\Sigma_Z^{-1}B+D_{1,n-1}^TD_{2,n-1}\\[2pt]
B^T\Sigma_Z^{-1}A +D^T_{2,n-1}D_{1,n-1}   & 1+B^T\Sigma_Z^{-1}B+D_{2,n-1}^TD_{2,n-1}
\end{bmatrix}.
\end{aligned}
\end{equation*}
Its inverse can be written as
\begin{equation*}
\hspace{-0.8cm}
\begin{aligned}
 (\Delta_{2,n-1})^{-1}&=\frac{1}{\det(\Delta_{2,n-1})}\begin{bmatrix}
    1+B^T\Sigma_Z^{-1}B+D_{2,n-1}^TD_{2,n-1}& -A^T\Sigma_Z^{-1}B-D_{1,n-1}^TD_{2,n-1}\\
    -B^T\Sigma_Z^{-1}A -D^T_{2,n-1}D_{1,n-1}&1+A^T\Sigma_Z^{-1}A +D_{1,n-1}^TD_{1,n-1}
 \end{bmatrix}\\
 &=\frac{1}{\det(\Delta_{2,n-1})}\left(\begin{bmatrix}1+D_{2,n-1}^TD_{2,n-1} & -D^T_{1,n-1}D_{2,n-1}\\ -D^T_{2,n-1}D_{1,n-1} & 1+D_{1,n-1}^TD_{1,n-1}\end{bmatrix} +\begin{bmatrix}B^T\Sigma_Z^{-1}B & -A^T\Sigma_Z^{-1}B \\ -B^T\Sigma_Z^{-1}A & A^T\Sigma_Z^{-1}A \end{bmatrix}              \right),
\end{aligned}
\end{equation*}
which, substituted into \eqref{MMSE-Vn-final} and then combined with \eqref{power-eqn-sk2}, gives
\begin{equation}\label{MMSE-Vn2}
\begin{aligned}
 \E[|X_n|^2]&= \frac{a_n^2+b_n^2+\|a_nD_{2,n-1}-b_nD_{1,n-1} \|^2+(a_nB^T-b_nA^T)\Sigma_Z^{-1}(a_nB-b_nA)}{\det(\Delta_{2,n-1})},
\end{aligned}
\end{equation}
where $\|\cdot\|$ denotes the Euclidean norm.

We are now ready to evaluate the limits of \eqref{pf-sk2-mutual-info} and (\ref{MMSE-Vn2}). Since $d_{1,n}$ and $d_{2,n}$ can be alternatively expressed as 
\begin{equation*}
\begin{aligned}
d_{1,n}=\begin{bmatrix}-\gamma_2 &-\gamma_1\end{bmatrix} \begin{bmatrix} c_1L(\gamma_1^{-1})\gamma_1^n\\[2pt]c_2L(\gamma_2^{-1})\gamma_2^n\end{bmatrix}, ~
d_{2,n}=\begin{bmatrix}1&1 \end{bmatrix}\begin{bmatrix} c_1L(\gamma_1^{-1})\gamma_1^n\\[2pt]c_2L(\gamma_2^{-1})\gamma_2^n\end{bmatrix},
\end{aligned}
\end{equation*}
it follows that
\begin{equation}\label{2norm-D1n}
\begin{aligned}
\|D_{1,n}\|^2&=D^T_{1,n}D_{1,n}     \\
 &=\sum_{k=p+1}^n\vert d_{1,k}\vert^2\\
&=\sum_{k=p+1}^n\begin{bmatrix}-\gamma_2&-\gamma_1\end{bmatrix}\begin{bmatrix}c_1L(\gamma_1^{-1})\gamma_1^k\\[2pt]c_2L(\gamma_2^{-1})\gamma_2^k\end{bmatrix}\begin{bmatrix}c_1L(\gamma_1^{-1})\gamma_1^k&c_2L(\gamma_2^{-1})\gamma_2^k\end{bmatrix}\begin{bmatrix}-\gamma_2\\[2pt]-\gamma_1\end{bmatrix} \\[2pt]
&=\sum_{k=p+1}^n \begin{bmatrix}\gamma_2&\gamma_1\end{bmatrix}\begin{bmatrix} c_1^2L^2(\gamma_1^{-1})\gamma_1^{2k} & c_1c_2L(\gamma_1^{-1})L(\gamma_2^{-1})\gamma_1^k \gamma_2^k\\[2pt]c_1c_2L(\gamma_1^{-1})L(\gamma_2^{-1})\gamma_1^k \gamma_2^k & c_2^2L^2(\gamma_2^{-1})\gamma_2^{2k}\end{bmatrix} \begin{bmatrix}\gamma_2\\[2pt]\gamma_1\end{bmatrix}\\[2pt]
&=\begin{bmatrix}\gamma_2&\gamma_1\end{bmatrix} 
\begin{bmatrix}
L_{11}\gamma_1^2(\gamma_1^{2n}-1) &L_{12}\gamma_1\gamma_2(\gamma_1^n\gamma_2^n-1)\\[2pt]L_{12}\gamma_1\gamma_2(\gamma_1^n\gamma_2^n-1) &L_{22}\gamma_2^2(\gamma_2^{2n}-1)
\end{bmatrix}
\begin{bmatrix}\gamma_2\\[2pt]\gamma_1\end{bmatrix}-R_1\\[2pt]
&=\begin{bmatrix}\gamma_2&\gamma_1\end{bmatrix}L_{n}^{(2)}\begin{bmatrix}\gamma_2\\[2pt]\gamma_1\end{bmatrix}-R_1,
\end{aligned}
\end{equation}
where we have defined $R_1=\sum_{k=1}^p |d_{1,k}|^2$ and
$$
L_{n}^{(2)}=\begin{bmatrix}
L_{11}\gamma_1^2(\gamma_1^{2n}-1) &L_{12}\gamma_1\gamma_2(\gamma_1^n\gamma_2^n-1)\\[2pt]L_{12}\gamma_1\gamma_2(\gamma_1^n\gamma_2^n-1) &L_{22}\gamma_2^2(\gamma_2^{2n}-1)
\end{bmatrix}.
$$
Next, set $R_2 = \sum_{k=1}^p |d_{2,k}|^2$ and $R_{12} = \sum_{k=1}^p d_{1,k} d_{2,k}$. Then, in complete analogy to the derivations for $D^T_{1,n}D_{1,n}$, we obtain
\begin{equation}\label{2norm-D2n}
\begin{aligned}
D_{2,n}^TD_{2,n}=\begin{bmatrix}1&1\end{bmatrix} L_n^{(2)}\begin{bmatrix}1\\[2pt]1\end{bmatrix}-R_2,~
D_{1,n}^TD_{2,n}=\begin{bmatrix}-\gamma_2&-\gamma_1\end{bmatrix} L_n^{(2)}\begin{bmatrix}1\\[2pt]1\end{bmatrix}-R_{12}.
\end{aligned}
\end{equation}
Hence, 
\begin{equation}\label{diff-D12n}
\begin{aligned}
   D_{1,n}^T&D_{1,n}D_{2,n}^TD_{2,n}-D_{1,n}^TD_{2,n}D_{2,n}^TD_{1,n}=\sum_{p+1\le i<j\le n}(d_{1,i}d_{2,j}-d_{1,j}d_{2,i})^2\\
   &=(\gamma_1-\gamma_2)^2c_1^2c_2^2\gamma_1^2\gamma_2^2L^2(\gamma_1^{-1})L^2(\gamma_2^{-1})\left(\frac{(\gamma_1^{2n}-1)(\gamma_2^{2n}-1)}{(\gamma_1^2-1)(\gamma_2^2-1)} - \frac{(\gamma_1^n\gamma_2^n-1)^2}{(\gamma_1\gamma_2-1)^2}  \right)\\[2pt]
   &~~~~-R_1 D^T_{2,n}D_{2,n}-R_2 D^T_{1,n}D_{1,n}+2R_{12}D^T_{1,n}D_{2,n}-R_1R_2+R^2_{12}\\
   &=\Delta \gamma_1^2\gamma_2^2(\gamma_1^{2n}\gamma_2^{2n}+1) + 2L^2_{12}\gamma_1^2\gamma_2^2(\gamma_1-\gamma_2)^2\gamma_1^n\gamma_2^n -\frac{(\gamma_1\gamma_2-1)^2\Delta \gamma_1^2\gamma_2^2}{(\gamma_1-\gamma_2)^2}(\gamma_1^{2n}+\gamma_2^{2n})\\[2pt]
   &~~~~-R_1 D^T_{2,n}D_{2,n}-R_2 D^T_{1,n}D_{1,n}+2R_{12}D^T_{1,n}D_{2,n}-R_1R_2+R^2_{12}.
\end{aligned}
\end{equation}
Furthermore, we have
\begin{equation}\label{det-sk2-1}
\hspace{-0.5cm}
\begin{aligned}
\det(\Delta_{2,n}) &= \det(I_2+\widetilde{D}_n^T\widetilde{\Sigma}_{Z,n}^{-1}\widetilde{D}_n)       \\
&= \det \begin{pmatrix}
   1+D_{1,n}^TD_{1,n} +A^T\Sigma_Z^{-1}A & D_{1,n}^TD_{2,n}+A^T\Sigma_Z^{-1}B\\[2pt] D_{2,n}^TD_{1,n} +B^T\Sigma_Z^{-1}A & 1+D_{2,n}^TD_{2,n}+B^T\Sigma_Z^{-1}B \end{pmatrix}\\
   &=(1+D_{1,n}^TD_{1,n})(1+D_{2,n}^TD_{2,n}) -D_{1,n}^TD_{2,n}D_{2,n}^TD_{1,n} +B^T\Sigma_Z^{-1}B(1+D^T_{1,n}D_{1,n}) \\[2pt]
   &~~~~+A^T\Sigma_Z^{-1}A(1+D^T_{2,n}D_{2,n}) +A^T\Sigma_Z^{-1}AB^T\Sigma_Z^{-1}B - 2A^T\Sigma_Z^{-1}BD^T_{1,n}D_{2,n}-(A^T\Sigma_Z^{-1}B)^2,
    \end{aligned}
\end{equation}
which, together with (\ref{2norm-D1n})--(\ref{diff-D12n}), immediately implies that
\begin{equation}\label{det-sk2}
\det(\Delta_{2,n})= \Delta \gamma_1^2\gamma_2^2\gamma_1^{2n}\gamma_2^{2n} +O\left((|\gamma_1|\vee|\gamma_2|)^{2n}\right),
\end{equation}
where we have adopted the notation $\phi\vee \psi\triangleq \max(\phi,\psi)$.

Consequently, it follows from (\ref{det-sk2}) that, upon taking the limit as $n\to\infty$ in \eqref{pf-sk2-mutual-info}, the mutual information rate achieved by the SK(2) coding scheme specified in equations (\ref{SK2-coding-input}) and (\ref{SK2-coding-message}) is given by
\begin{equation}\label{mutual-info-sk2-proof}
\begin{aligned}
            \bar{I}(V;Y)=\lim_{n\to\infty}\frac{1}{n}I(V_1^n;Y_1^n)= \log|\gamma_1 \gamma_2|.
\end{aligned}
\end{equation}

On the other hand, we now examine the limiting behavior of \eqref{MMSE-Vn2} as \(n \to \infty\). To this end, observe that
$$
a_nd_{2,k}-b_nd_{1,k}=\begin{bmatrix}\gamma_2^{n-1}&\gamma_1^{n-1}\end{bmatrix} \begin{bmatrix}
   c_1L(\gamma_1^{-1})\gamma_1^k \\[2pt]c_2L(\gamma_2^{-1})\gamma_2^k
\end{bmatrix},
$$
which implies that 
\begin{equation}\label{aD-bD}
\begin{aligned}
\|&a_nD_{2,n-1}-b_nD_{1,n-1} \|^2= \begin{bmatrix}
      \gamma_2^{n-1}&\gamma_1^{n-1}
\end{bmatrix}
L_{n-1}^{(2)}
\begin{bmatrix}
      \gamma_2^{n-1}\\[2pt]\gamma_1^{n-1}
\end{bmatrix} - \begin{bmatrix}\gamma_2^{n-1}&\gamma_1^{n-1}\end{bmatrix}L_p^{(2)}\begin{bmatrix}\gamma_2^{n-1}\\[2pt]\gamma_1^{n-1}\end{bmatrix}\\
&=L_{11}\gamma_1^2\gamma_2^{2(n-1)}(\gamma_1^{2(n-1)}-1) +L_{22}\gamma_2^2\gamma_1^{2(n-1)}(\gamma_2^{2(n-1)}-1)+2L_{12}\gamma_1^{n}\gamma_2^{n}(\gamma_1^{n-1}\gamma_2^{n-1}-1) \\[2pt]
&~~~~-(L_p^{(2)})_{11}\gamma_2^{2(n-1)}-(L_p^{(2)})_{22}\gamma_1^{2(n-1)}-2(L_p^{(2)})_{12}\gamma^{n-1}_1\gamma^{n-1}_2.
\end{aligned}
\end{equation}
Similarly, for $1\le k\le p$, we have
\begin{equation*}
\begin{aligned}
     a_nb_k -b_na_k = c_1c_2(\gamma_1-\gamma_2)\begin{bmatrix}-\gamma_2^n &\gamma_1^n\end{bmatrix} \begin{bmatrix}\gamma_1^k\\[2pt]\gamma_2^k\end{bmatrix}.
\end{aligned}  
\end{equation*}
Consequently,
\begin{equation}\label{aB-bA}
\begin{aligned}
     (a_nB^T-b_nA^T)\Sigma_Z^{-1}(a_nB-b_nA)&=c_1^2c_2^2(\gamma_1-\gamma_2)^2\begin{bmatrix}-\gamma_2^n &\gamma_1^n\end{bmatrix}T\begin{bmatrix}-\gamma_2^n \\[2pt]\gamma_1^n\end{bmatrix}\\
     &=c_1^2c_2^2(\gamma_1-\gamma_2)^2\left(T_{11}\gamma_2^{2n}+T_{22}\gamma_1^{2n}-(T_{12}+T_{21})\gamma_1^n\gamma_2^n\right),
\end{aligned}  
\end{equation}
where $T$ is the $2 \times 2$ matrix defined by
\[
T = \sum_{i=1}^p \sum_{j=1}^p
\begin{bmatrix} \gamma_1^i \\[2pt] \gamma_2^i \end{bmatrix}
(\Sigma_Z^{-1})_{ij}
\begin{bmatrix} \gamma_1^j & \gamma_2^j \end{bmatrix}.
\]

Taking the limit $n \to \infty$ in \eqref{MMSE-Vn2}, we obtain
\begin{equation}\label{power-eqn-sk2-proof}
   \begin{aligned}
      \lim_{n\to\infty}\E[|X_n|^2]
      &=\lim_{n\to\infty}\frac{a_n^2+b_n^2+\|a_nD_{2,n-1}-b_nD_{1,n-1}\|^2 +(a_nB^T-b_nA^T)\Sigma_Z^{-1}(a_nB-b_nA)}{\det(\Delta_{2,n-1})}\\
      &\overset{(f)}{=}\frac{L_{11}\gamma_1^2 +L_{22}\gamma_2^2+2L_{12}\gamma_1\gamma_2}{\Delta \gamma_1^2\gamma_2^2}\\
      &=\frac{1}{\Delta}\left(\frac{L_{11}}{\gamma_2^2}+\frac{L_{22}}{\gamma_1^2}+\frac{2L_{12}}{\gamma_1\gamma_2}\right)\\
      &= P_2(\gamma_1,\gamma_2).
   \end{aligned}
\end{equation}
where $(f)$ follows from (\ref{det-sk2}), (\ref{aD-bD}) and \eqref{aB-bA}. The limit \eqref{power-eqn-sk2-proof}, together with \eqref{power-constraint-sk1}, immediately implies that
\begin{equation}\label{power-eqn-sk2-proof-2}
\lim_{n\to\infty} \frac{1}{n} \sum_{i=1}^n \mathbb{E}[X_i^2]=P_2(\gamma_1,\gamma_2)\le P.
\end{equation}
Remark~\ref{remark-exact-Power:label} permits the equality constraint in \eqref{I-SK2-thm-constraints}. Therefore, for distinct admissible roots,  it follows from \eqref{mutual-info-sk2-proof} and \eqref{power-eqn-sk2-proof-2} that the desired results in \eqref{I-SK2-thm} and \eqref{power-eqn-sk2-thm} hold, completing the proof for Case A.

We now establish the result for Case B. Recall that in this case we have $\gamma_1=\gamma_2=\gamma\in \mathbb{R}$ with $|\gamma|>1$, which yields $a_n=(2-n)\gamma^{n-1}$ and $b_n=(n-1)\gamma^{n-2}$ as in (\ref{Vn-case-b}). Following the same construction as in \eqref{Y-tlta-re}, we define \(\{\widetilde{Y}\}\) accordingly, which leads to
\begin{equation*}\label{d12-case-c}
\left\{
\begin{aligned}
d_{1,n} &=\left[(2-n)L(\gamma^{-1})+L^\prime(\gamma^{-1})\gamma^{-1} \right]  \gamma^{n-1}\\[2pt]
d_{2,n} &= \left[(n-1)L(\gamma^{-1}) -L^\prime(\gamma^{-1})\gamma^{-1}   \right]\gamma^{n-2},
\end{aligned}
\right.
\end{equation*}
where $L^\prime(\gamma^{-1})$ denotes the derivative of $L(z)$ evaluated at $z = \gamma^{-1}$. 

Consequently, the scalar products $D^T_{1,n}D_{1,n}$, $D^T_{2,n}D_{2,n}$ and $D^T_{1,n}D_{2,n}$ are computed as follows:
\begin{equation}\label{D12n-case-c}
\begin{aligned}
D^T_{1,n}D_{1,n}&= L^2(\gamma^{-1})\left(\gamma^2S^{(2)}_{n-1}+1\right) -2L(\gamma^{-1})L^\prime(\gamma^{-1})\left(\gamma S^{(1)}_{n-1} -\gamma^{-1}\right)-R_1\\
&~~~~+\frac{(L^\prime(\gamma^{-1}))^2\gamma^{-2}}{\gamma^2-1}(\gamma^{2n}-1),\\
D^T_{2,n}D_{2,n}&= L^2(\gamma^{-1})\gamma^{-2}S_n^{(2)} -2L(\gamma^{-1})L^\prime(\gamma^{-1})\gamma^{-3}S_n^{(1)} +\frac{(L^\prime(\gamma^{-1}))^2\gamma^{-4}}{\gamma^2-1}(\gamma^{2n}-1)-R_2,\\
D^T_{1,n}D_{2,n}&= L(\gamma^{-1})\left(L^\prime(\gamma^{-1})+L^\prime(\gamma^{-1})\gamma^{-2}-L(\gamma^{-1})\gamma\right)S_{n-1}^{(1)}-R_{12}\\
&~~~~+L(\gamma^{-1})L^\prime(\gamma^{-1})\gamma^{-4}\gamma^{2n}(n-1) - L^2(\gamma^{-1})\gamma S^{(2)}_{n-1} \\
&~~~~-\frac{(L^\prime(\gamma^{-1}))^2\gamma^{-3}}{\gamma^2-1}(\gamma^{2n}-1) - L(\gamma^{-1})L^\prime(\gamma^{-1})\gamma^{-2},
\end{aligned}
\end{equation}
where
\begin{equation}\label{Sn1-case-c}
\begin{aligned}
S^{(1)}_n \triangleq \sum_{k=1}^{n-1}k\gamma^{2k}=\frac{1}{\gamma^2-1}n\gamma^{2n} - \frac{\gamma^4}{(\gamma^2-1)^2}\gamma^{2(n-1)} + \frac{\gamma^2}{(\gamma^2-1)^2}
\end{aligned}
\end{equation}
and
\begin{equation}\label{Sn2-case-c}
   \begin{aligned}
      S^{(2)}_n &\triangleq \sum_{k=1}^{n-1}k^2\gamma^{2k}=\frac{1}{\gamma^2-1}n^2\gamma^{2n} - \frac{2\gamma^2}{(\gamma^2-1)^2}n\gamma^{2n} + \frac{\gamma^4(\gamma^2+1)}{(\gamma^2-1)^3}\gamma^{2(n-1)}-\frac{(\gamma^2+1)\gamma^2}{(\gamma^2-1)^3}.
   \end{aligned}
\end{equation}
Proceeding analogously to the derivation of (\ref{det-sk2-1}) and (\ref{det-sk2}), it follows from \eqref{D12n-case-c}, \eqref{Sn1-case-c} and \eqref{Sn2-case-c} that 
\begin{equation}\label{det-sk2-case-c}
   \begin{aligned}
   \det(\Delta_{2,n})&=\frac{L^4(\gamma^{-1})}{(\gamma^2-1)^4}\gamma^{4n} + O(n^2|\gamma|^{2n}).
   \end{aligned}
\end{equation}
Applying this to \eqref{pf-sk2-mutual-info} and subsequently taking the limit as $n\to\infty$ leads to
\begin{equation}\label{pf-sk2-mutual-info-casec}
\begin{aligned}
\bar{I}(V;Y)&=\lim_{n\to\infty}\frac{1}{n}I(V_1^n;Y_1^n)= 2\log |\gamma|.
\end{aligned}
\end{equation}

We next evaluate the limit of the power usage as in \eqref{MMSE-Vn2}. To this end, observe that
$$
a_n d_{2,k} - b_nd_{1,k} = \gamma^{n-3}(H^{(2)}_n)^T\begin{bmatrix}
   k\\1
\end{bmatrix}
\gamma^k,
$$
where 
$$
H^{(2)}_n \triangleq \begin{bmatrix}L(\gamma^{-1}) & -\left(nL(\gamma^{-1}) +\gamma^{-1}L^\prime(\gamma^{-1})\right)\end{bmatrix}^T.
$$
This implies that 
\begin{equation}\label{aD2-bD1-case-c}
\hspace{-0.3cm}
   \begin{aligned}
      \|a_nD_{2,n-1} - b_nD_{1,n-1}  \|^2 &= \sum_{k=p+1}^{n-1}(a_nd_{2,k} - b_nd_{1,k})^2\\
      &=\gamma^{2(n-3)}\sum_{k=p+1}^{n-1}(H^{(2)}_n)^T\begin{bmatrix}
         k\\ 1
      \end{bmatrix}
      \begin{bmatrix}
         k&1
      \end{bmatrix}
      H^{(2)}_n \gamma^{2k}\\[2pt]
      &= (H^{(2)}_n)^T\begin{bmatrix}
         \sum_{k=p+1}^{n-1} k^2\gamma^{2k} & \sum_{k=p+1}^{n-1}k\gamma^{2k}\\[2pt]
         \sum_{k=p+1}^{n-1}k\gamma^{2k}& \sum_{k=p+1}^{n-1}\gamma^{2k}
      \end{bmatrix}
      H^{(2)}_n \gamma^{2(n-3)}\\[2pt]
      &= \gamma^{2(n-3)}(H^{(2)}_n)^T\begin{bmatrix}
         S_n^{(2)}-\sum_{k=1}^pk^2\gamma^{2k} & S_n^{(1)}-\sum_{k=1}^pk\gamma^{2k}\\[2pt]
         S_n^{(1)}-\sum_{k=1}^pk\gamma^{2k}& \frac{\gamma^{2n}-\gamma^{2(p+1)}}{\gamma^2-1}
      \end{bmatrix}
      H^{(2)}_n.
   \end{aligned}
\end{equation}

Then, taking the limit as $n\to\infty$ in \eqref{MMSE-Vn2} gives
\begin{equation}\label{power-eqn-case-c}
   \hspace{-0.5cm}
   \begin{aligned}
      \lim_{n\to\infty}\E[|X_n|^2] 
      &= \lim_{n\to\infty} \frac{a_n^2 +b_n^2 +\|a_nD_{2,n-1} - b_nD_{1,n-1}  \|^2 +(a_nB^T-b_nA^T)\Sigma_Z^{-1}(a_nB-b_nA)}{\det(\Delta_{2,n-1})}\\
      &\overset{(g)}{=} \frac{(\gamma^2-1)^3(L^\prime(\gamma^{-1}))^2}{\gamma^4 L^4(\gamma^{-1})} + \frac{2(\gamma^2-1)^2}{\gamma}\frac{L^\prime(\gamma^{-1})}{L^3(\gamma^{-1})} +\frac{\gamma^4-1}{L^2(\gamma^{-1})},
   \end{aligned}
\end{equation}
where $(g)$ follows from the definitions of $a_n$ and $b_n$ together with (\ref{det-sk2-case-c}) and (\ref{aD2-bD1-case-c}).

It remains to show that the distinct-root expression in \eqref{power-eqn-sk2-thm} converges to the right-hand side of \eqref{power-eqn-case-c} as $\gamma_1\to\gamma$ and $\gamma_2=\gamma\in\mathbb{R}$. More precisely,
\begin{equation}\label{power-eqn-equivalence-case-c}
\hspace*{-0.3cm}
\begin{aligned}
\lim_{\substack{\gamma_1\to\gamma \\ \gamma_2=\gamma\in\mathbb{R}}} \frac{1}{\Delta}\left(\frac{L_{11}}{\gamma_2^2} + \frac{L_{22}}{\gamma_1^2} + \frac{2L_{12}}{\gamma_1\gamma_2}\right) &= \frac{(\gamma^2-1)^3(L^\prime(\gamma^{-1}))^2}{\gamma^4 L^4(\gamma^{-1})} + \frac{2(\gamma^2-1)^2}{\gamma}\frac{L^\prime(\gamma^{-1})}{L^3(\gamma^{-1})}+\frac{\gamma^4-1}{L^2(\gamma^{-1})}.
\end{aligned}
\end{equation}
Together with \eqref{pf-sk2-mutual-info-casec} and \eqref{power-eqn-case-c}, this immediately establishes the desired results as in \eqref{I-SK2-thm} and \eqref{power-eqn-sk2-thm} for Case B. The proof of the identity (\ref{power-eqn-equivalence-case-c}) is deferred to Appendix \ref{sec:proof_of_ref_power_eqn_equivalence_case_c}.
\end{proof}

\begin{lem}\label{lemma2:label}
Consider the SK(2) coding scheme \eqref{SK2-coding-input}--\eqref{SK2-coding-message} over the stationary AR($p$) Gaussian channel \eqref{channel-arp}, under the average-power constraint \eqref{power-constraint-sk1} with $P>0$. If at most one characteristic root is unstable, with roots counted with algebraic multiplicity, then 
$$
\bar{I}(V;Y)\le C_\mathrm{SK1}(P).
$$
\end{lem}

\begin{proof}
The proof is deferred to Appendix~\ref{sec-app:reduction_lemma}.
\end{proof}


The preceding results cover all characteristic-root configurations of the full SK(2) class. We now show that optimizing over admissible pairs already gives the full SK(2) capacity.

\begin{co}\label{main-col-sk2:label}
Consider the stationary AR($p$) Gaussian channel \eqref{channel-arp} with noise power spectral density \eqref{sdf-arp}. For every $P>0$,
$$
C_\mathrm{SK2}(P)=C^g_\mathrm{SK2}(P)\ge C_\mathrm{SK1}(P).
$$
\end{co}

\begin{proof}
Let 
$$
P_1(\gamma) = (\gamma^2 - 1)H_Z^2(\gamma^{-1}), \qquad \gamma \in \mathbb{R}, \quad |\gamma| > 1.
$$
By Theorem~\ref{ISK1-thm}, choose an optimizer $\gamma^\ast\in \mathbb{R}$ such that 
\begin{equation}
    P_1(\gamma^\ast) = P, \qquad C_{\mathrm{SK}1}(P) = \log|\gamma^\ast|.
\end{equation}
For distinct real numbers $x,y$ satisfying $|x|\wedge |y|>1$, the identity
$$
(xy - 1)^2 - (x^2 - 1)(y^2 - 1) = (x - y)^2
$$
and \eqref{power-eqn-sk2-thm}--\eqref{main-thm-notations} yield
\begin{equation}\label{eq:P2-arp-co}
P_2(x,y) = \frac{1}{(x - y)^2} \left[ \frac{(y^2 - 1)(xy - 1)^2}{L^2_Z(y^{-1})} + \frac{(x^2 - 1)(xy - 1)^2}{L^2_Z(x^{-1})} \right.
\left. - \frac{2(x^2 - 1)(y^2 - 1)(xy - 1)}{L_Z(x^{-1})L_Z(y^{-1})} \right].
\end{equation}
Since $L_Z$ has no zeros on the closed unit disk, \eqref{eq:P2-arp-co} shows that $P_2(x,y)$ extends to a smooth real-valued function near $(s,\gamma^\ast)$ for either $s\in\{-1,1\}$. In particular, we have
\begin{equation}\label{eq:P2-arp-co-P1}
    P_2(s, y) \triangleq \lim_{x\to s}P_2(x, y) = \frac{(y^2 - 1)(sy - 1)^2}{L_Z^2(y^{-1})(s - y)^2} = P_1(y).
\end{equation}
Furthermore, the repeated-root formula \eqref{power-eqn-case-c} yields
\begin{equation}\label{eq:pf-P2-P1-d}
P_2(\gamma, \gamma) = P_1(\gamma) + \frac{(\gamma^2 - 1)L_Z^2(\gamma^{-1})}{4} [P_1^\prime(\gamma)]^2.
\end{equation}
If $P^\prime_1(\gamma^\ast)\neq 0$, choose real $\gamma_n$ such that $\lim_{n\to\infty}\gamma_n=\gamma^\ast$, $|\gamma_n|>1$ and $P_1(\gamma_n)<P$. For each $n$, by \eqref{eq:P2-arp-co-P1}, there exists $s_n$ such that
$$
1 < s_n < 1 + \frac{1}{n}, \qquad P_2(s_n, \gamma_n) < P.
$$
For all sufficiently large $n$, the pairs $(s_n, \gamma_n)$ are real, distinct, and admissible. Then, it follows from Theorem~\ref{main-theorem-sk2} and Remark~\ref{remark-exact-Power:label} that
\begin{equation}
\begin{aligned}
C_{\mathrm{SK}2}^g(P) \geq \lim_{n\to\infty}\log|s_n\gamma_n| = \log|\gamma^\ast| = C_{\mathrm{SK}1}(P).
\end{aligned}
\end{equation}
On the other hand, if $P^\prime_1(\gamma^\ast)= 0$, by \eqref{eq:pf-P2-P1-d}, we have
$$
P_2(\gamma^\ast,\gamma^\ast)=P_1(\gamma^\ast)=P.
$$
Consequently, the repeated-root pair $(\gamma^\ast,\gamma^\ast)$ is admissible, and by Theorem~\ref{main-theorem-sk2}, we have that the information rate $R_2(\gamma^\ast,\gamma^\ast)=2 \log |\gamma^\ast|$. Hence
$$
C_{\mathrm{SK}2}^g(P) \geq 2\log|\gamma^\ast| = 2C_{\mathrm{SK}1}(P) > C_{\mathrm{SK}1}(P).
$$
The two cases establish $C^g_\mathrm{SK2}(P)\ge C_\mathrm{SK1}(P)$.

Finally, note that every feasible SK(2) coding scheme either has an admissible characteristic pair or has at most one unstable root. Thus, the supremum over the former class is $C^g_\mathrm{SK2}(P)$, whereas Lemma~\ref{lemma2:label} bounds every mutual information rate in the latter class by $C_\mathrm{SK1}(P)$. Consequently, 
$$
C^g_{\mathrm{SK}2}(P)\leq C_{\mathrm{SK}2}(P) \leq \max\{C_{\mathrm{SK}2}^g(P), C_{\mathrm{SK}1}(P)\} = C_{\mathrm{SK}2}^g(P).
$$
Therefore, $C_{\mathrm{SK}2}(P) = C_{\mathrm{SK}2}^g(P) \geq C_{\mathrm{SK}1}(P)$, as desired.
\end{proof}

\begin{co}\label{co-twice-sk2:label}
Consider the stationary AR($p$) Gaussian channel \eqref{channel-arp} having power spectral density \eqref{sdf-arp} and subject to the average-power constraint \eqref{power-constraint-sk1}, where $P>0$. Suppose that an optimizer $\gamma_\ast$ in Theorem~\ref{ISK1-thm} satisfies
$$
P_1^\prime(\gamma_\ast)=0,
$$
where $P_1(\gamma) \triangleq (\gamma^2 - 1)/L^2_Z(\gamma^{-1})$ for $\gamma\in\mathbb{R}$ and $|\gamma|>1$. Then, the admissible repeated-root pair $(\gamma_\ast,\gamma_\ast)$ has asymptotic input power $P$ and information rate $2C_\mathrm{SK1}(P)$. In particular,
$$
C_{\mathrm{SK}2}^g(P) \geq 2C_{\mathrm{SK}1}(P) > C_{\mathrm{SK}1}(P).
$$
\end{co}
\begin{proof}
It follows from \eqref{eq:pf-P2-P1-d} that $P_2(\gamma_\ast,\gamma_\ast)=P_1(\gamma_\ast)=P$. Since $|\gamma_\ast|>1$, the pair $(\gamma_\ast,\gamma_\ast)$ is admissible for the SK(2) coding scheme. By Theorem~\ref{main-theorem-sk2}, we have
$$
C_{\mathrm{SK}2}^g(P) \geq 2\log|\gamma_\ast| = 2C_{\mathrm{SK}1}(P) > C_{\mathrm{SK}1}(P),
$$
where the last inequality follows from $\log |\gamma_\ast|>0$, proving the result.
\end{proof}

\subsection{Capacity-Achieving Cases}\label{subsec:capacity_achieving_cases}
We next provide three cases in which the SK(2) scheme achieves feedback capacity: the AWGN channel \eqref{AWGN-channel}, stationary AR(1) Gaussian channels, and an interleaved AR(1) subclass of stationary AR(2) Gaussian channels.

We begin with the flat noise spectrum
\begin{equation}\label{AWGN-SDF}
   S_Z(e^{i\theta})\equiv 1.
\end{equation}
This spectral density corresponds to a stationary noise process of the form $Z_i = W_i,~i\ge 1$, where $\{W_i\}_{i=1}^{\infty}$ is a zero-mean, unit-variance white Gaussian process. This corresponds precisely to the AWGN channel (\ref{AWGN-channel}). 

When $b=0$ and $|a|>1$, the asymptotic rate and power coincide with those of the SK(1) coding scheme with $\gamma=a$, as shown in Appendix~\ref{sec-app:reduction_lemma}. Thus, the SK(2) class includes capacity-achieving schemes for this channel by Remark~\ref{remark1:label}. Within the genuine SK(2) class, the following corollary demonstrates that, for this channel, the optimal pair $(\gamma_1,\gamma_2)$ in the optimization problem (\ref{I-SK2-thm})--(\ref{power-eqn-sk2-thm}) satisfies $|\gamma_1\gamma_2|=\sqrt{P+1}$.  Consequently, the SK(2) coding scheme yields a broader family of optimal coding schemes for the AWGN channel \eqref{AWGN-channel}.

\begin{co}\label{co-awgn-sk2:label}
Consider the noise process $\{Z_i\}_{i=1}^\infty$ with the power spectral density (\ref{AWGN-SDF}). Then, the feedback capacity of the AWGN channel $Y_i = X_i +Z_i,~i=1,2,\dots$, subject to average power constraint $P$, can be achieved via the SK(2) coding scheme defined in \eqref{SK2-coding-input} and \eqref{SK2-coding-message} with any admissible pair $(\gamma_1,\gamma_2)$ satisfying $|\gamma_1\gamma_2|=\sqrt{P+1}$.
\end{co}

\begin{proof}
For the AWGN channel \eqref{AWGN-channel}, $L_Z(e^{i\theta})\equiv 1$. Applying Theorem~\ref{main-theorem-sk2}, the optimization problem formulated in (\ref{I-SK2-thm}), (\ref{power-eqn-sk2-thm}) and (\ref{main-thm-notations}) can be simplified. More precisely, every admissible pair $(\gamma_1,\gamma_2)$ has
\begin{equation}\label{awgn-sk2-power-eqn}
\begin{aligned}
P_2(\gamma_1,\gamma_2)=\gamma^2_1 \gamma^2_2-1=|\gamma_1\gamma_2|^2-1,
\end{aligned}
\end{equation}
where the last equality follows from $\gamma_1\gamma_2\in\mathbb{R}$. As a result, the power constraint $P_2(\gamma_1,\gamma_2)\le P$ implies that
$$
R_2(\gamma_1,\gamma_2)\le \frac{1}{2}\log (1+P),
$$
with equality if and only if $|\gamma_1\gamma_2|=\sqrt{1+P}$. Such admissible pairs exist; for example, $\gamma_1=\gamma_2=(1+P)^{1/4}$. Hence, every admissible pair satisfying $|\gamma_1\gamma_2|=\sqrt{1+P}$ achieves the feedback capacity of the AWGN channel.

\end{proof}

We next consider the nonwhite stationary AR(1) Gaussian channel 
\[
Y_i=X_i+Z_i,\quad i\ge1,
\]
with noise power spectral density
\begin{equation}\label{ar1-sdf}
S_Z(e^{i\theta})=\frac{1}{|1+\beta_1e^{i\theta}|^2},\qquad 0<|\beta_1|<1.
\end{equation}
This spectral density corresponds to the stationary noise process
$$
Z_i +\beta_1 Z_{i-1} = W_i,\quad i\in\mathbb{Z}.
$$ 
By \cite[Theorem 5.2]{su2024feedback}, the feedback capacity of this channel under average power constraint $P$ is
\begin{equation}\label{CFB-AR1}
C_\mathrm{FB}(P) = -\log x_0,
\end{equation}
where $x_0$ is the unique positive root of the fourth-order polynomial equation
\begin{equation}\label{Power-Eqn-AR1}
Px^2 = \frac{1-x^2}{(1+ |\beta_1| x)^2}.
\end{equation}
As noted in Remark~\ref{remark1:label}, the SK(1) coding scheme defined in (\ref{SK1-coding-X}) and (\ref{SK1-coding-message}) is optimal for this channel; specifically,
\[
   C_\mathrm{SK1}(P)=C_\mathrm{FB}(P).
\]
Hence, by Corollary~\ref{main-col-sk2:label},
\begin{equation}\label{CSK2-AR1_equal}
   C_\mathrm{SK2}(P)=C_\mathrm{SK2}^g(P)=C_\mathrm{FB}(P).
\end{equation}
Thus, the SK(2) coding scheme with $b=0$ is optimal for the stationary AR(1) Gaussian channel. For nonwhite AR(1) noise, the following proposition shows that no fixed admissible pair attains this genuine supremum in \eqref{I-SK2-thm} and constructs power-feasible pairs whose rates approach it.

\begin{pr}\label{example-ar1:label}
Consider the stationary AR(1) Gaussian channel having noise power spectral density \eqref{ar1-sdf} and subject to the average-power constraint \eqref{power-constraint-sk1}, where $P>0$. For every admissible pair $(\gamma_1,\gamma_2)$ satisfying $P_2(\gamma_1,\gamma_2)\le P$, it holds that
\begin{equation}\label{pr-equ-ptp}
R_2(\gamma_1,\gamma_2)<C_\mathrm{FB}(P).
\end{equation}
Nevertheless,
\begin{equation}\label{pr-equ-sup}
C^g_\mathrm{SK2}(P)=C_\mathrm{SK1}(P)=C_\mathrm{FB}(P).
\end{equation}
The supremum in \eqref{I-SK2-thm} is not attained, but it can be approached by admissible pairs satisfying $P_2(\gamma_1,\gamma_2)= P$.
\end{pr}

\begin{proof}
For notational simplicity, let $\rho=|\beta_1|\in(0,1)$, $\sigma=\operatorname{sign}(\beta_1)$, $q=\gamma_1\gamma_2$ and $t=\sigma(\gamma_1+\gamma_2)$. For distinct roots, substituting $L_Z(z)=1+\beta_1 z$ into \eqref{power-eqn-sk2-thm}--\eqref{main-thm-notations} and using
\begin{equation*}
(q - 1)^2 - (\gamma_1^2 - 1)(\gamma_2^2 - 1) = (\gamma_1 - \gamma_2)^2,
\end{equation*}
we establish 
\begin{equation}\label{pr-ar1-power-eqn}
P_2(\gamma_1, \gamma_2) = \Pi_\rho(t, q),
\end{equation}
where 
$$
\Pi_\rho(t, q) \triangleq \frac{(q - 1) \left[ \rho^2(q - 1)t^2 + 2\rho q^2t + (q + 1)(q^2 + \rho^2) \right]}{(q + \rho t + \rho^2)^2}.
$$
It easily follows from \eqref{power-eqn-case-c}--\eqref{power-eqn-equivalence-case-c} that \eqref{pr-ar1-power-eqn} also holds for repeated roots by continuity. Clearly, since $|\gamma_1|\wedge |\gamma_2| > 1 > \rho$, it holds that
$$
q + \rho t + \rho^2 = (\sigma\gamma_1 + \rho)(\sigma\gamma_2 + \rho)\neq 0.
$$
Define 
$$
F_\rho(r) \triangleq \frac{r^2(r^2 - 1)}{(r + \rho)^2}, \qquad r > 1.
$$
Let $r_P=1/x_0$ and it then follows from \eqref{CFB-AR1}--\eqref{Power-Eqn-AR1} that
\begin{equation}\label{pf-ar1-F-rho-P}
F_\rho(r_P)=P,\qquad C_\mathrm{FB}(P)=\log r_P.
\end{equation}
Moreover, $F_\rho(r)$ is strictly increasing, since 
$$
\frac{F'_\rho(r)}{F_\rho(r)} = \frac{2\rho}{r(r + \rho)} + \frac{2r}{r^2 - 1} > 0.
$$
We claim that for every fixed admissible pair $(\gamma_1,\gamma_2)$, it holds that
\begin{equation}\label{pf-ar1-claim}
P_2(\gamma_1, \gamma_2) > F_\rho(|\gamma_1\gamma_2|).
\end{equation}
To see this, let $r=|q|>1$. The real-coefficient conditions in \eqref{def-admissible-pair} imply that $\gamma_1$ and $\gamma_2$ are either both real or complex conjugates. In either case,
$$
(q + 1)^2 - t^2 = (\gamma_1^2 - 1)(\gamma_2^2 - 1) > 0.
$$
Consequently, we have the following two cases:
$$
\begin{cases}
|t| < r + 1, & q = r > 1, \\
|t| < r - 1, & q = -r < -1.
\end{cases}
$$

\textbf{Case A}: $q = r > 1$. In this case, we have
\[
\Pi_\rho(t, r) - F_\rho(r) = \frac{\rho(r - 1)(r + 1 - t) \left[ 2r^3(1 - \rho)^2 + H_\rho(r)(t + r + 1) \right]}{(r + \rho)^2(r + \rho t + \rho^2)^2},
\]
where
$$
H_\rho(r) \triangleq \rho \left[ r + 1 + (1 - \rho)(r - 1)(\rho + 2r + 1) \right] > 0.
$$
Since $|t|<r+1$, it holds that
\begin{equation}\label{pf-ar1-case-a}
\Pi_\rho(t, r) > F_\rho(r).
\end{equation}

\textbf{Case B}: $q = -r < -1$. In this case, we can establish
\[
\Pi_\rho(t, -r) - F_\rho(r) = \frac{\rho(r + 1)(r - 1 - t) K_\rho(r,t)}{(r + \rho)^2(r - \rho t - \rho^2)^2},
\]
where 
\begin{equation*}
\begin{aligned}
\hspace*{-0.8cm}
K_\rho(r,t)\triangleq 2(r + \rho) \left[ (1 - \rho^2)r^2 + \rho r + \rho^2 \right] +(r - 1 - t)\left[\rho^3(r + 1) + 2\rho^2 r(r + 1) + 2\rho r^2\right]>0.
\end{aligned}
\end{equation*}
Clearly, we also have
\begin{equation}\label{pf-ar1-case-b}
\Pi_\rho(t, -r) > F_\rho(r).
\end{equation}
Combining \eqref{pf-ar1-case-a} in Case A and \eqref{pf-ar1-case-b} in Case B, we have shown that for every fixed admissible pair $(\gamma_1,\gamma_2)$, it holds that
\begin{equation}\label{pf-ar1-case-a-b}
P_2(\gamma_1,\gamma_2)>F_\rho(|\gamma_1\gamma_2|),
\end{equation}
which, together with the average power constraint $P_2(\gamma_1,\gamma_2)\le P$, immediately implies that
$$
F_\rho(|\gamma_1\gamma_2|)< P.
$$
It then follows from \eqref{pf-ar1-F-rho-P} and the strict monotonicity of $F_\rho$ that $|\gamma_1\gamma_2|<r_P$. Therefore, 
$$
R_2(\gamma_1,\gamma_2) = \log |\gamma_1\gamma_2| < \log r_P = C_{\mathrm{FB}}(P),
$$
which proves \eqref{pr-equ-ptp}.

It remains to prove \eqref{pr-equ-sup}. To achieve this, it can be seen clearly from \eqref{pf-ar1-claim}--\eqref{pf-ar1-case-a-b} that for $r>1$
\begin{equation}\label{pr-ar1-sup}
\Pi_\rho(r+1,r)=F_\rho(r),
\end{equation}
which corresponds to pairs on the boundary of the set of admissible pairs $(\gamma_1,\gamma_2)$, i.e., $(1,r)$ or $(-1,-r)$. We choose $1<r_n<r_P$ satisfying 
\begin{equation}\label{pf-ar1-lim-rP}
\lim_{n\to\infty}r_n =r_P.
\end{equation}
For each $r_n$, $\Pi_\rho(t,r_n)$ is continuous in a neighborhood of $t=r_n+1$. Thus, there exists $\delta_n>0$ such that for any $r_n+1-\delta_n<t\le r_n+1$, we always have
$$
\Pi_\rho(t,r_n)<P.
$$
We then choose $t_n$ such that $\max\{2\sqrt{r_n}, r_n+1-\delta_n\}<t_n<r_n+1$. Clearly, $\Pi_\rho(t_n,r_n)<P$. Moreover, consider the two roots $\widetilde{\gamma}_{1,n}$ and $\widetilde{\gamma}_{2,n}$ of the polynomial equation (in $z$) $z^2-t_nz+r_n=0$. It is straightforward to see that $\widetilde{\gamma}_{1,n}$ and $\widetilde{\gamma}_{2,n}$ are two distinct real roots satisfying $ \widetilde\gamma_{1,n} \wedge \widetilde\gamma_{2,n} >1$. Then, define $\gamma_{i,n}=\sigma\widetilde{\gamma}_{i,n}$ for $i=1,2$. It holds that
$$
|\gamma_{1,n}|\wedge |\gamma_{2,n}| >1, \qquad \gamma_{1,n} \gamma_{2,n}=r_n,\qquad \gamma_{1,n}+\gamma_{2,n}=\sigma t_n.
$$
Thus, $(\gamma_{1,n},\gamma_{2,n})$ is an admissible pair for SK(2) coding scheme corresponding to the characteristic equation in \eqref{sk2-char-eqn} with $b=-r_n\neq 0$. Moreover,
$$
P_2(\gamma_{1,n},\gamma_{2,n}) = \Pi_\rho(t_n,r_n) <P.
$$
By Theorem~\ref{main-theorem-sk2} and Remark~\ref{remark-exact-Power:label}, for every $n$,
$$
C^g_\mathrm{SK2}(P)\ge \log |\gamma_{1,n}\gamma_{2,n}|=\log r_n.
$$
Letting $n\to\infty$, by \eqref{pf-ar1-F-rho-P} and \eqref{pf-ar1-lim-rP}, we obtain
$$
C^g_\mathrm{SK2}(P)\ge \lim_{n\to\infty}\log r_n=\log r_P =C_\mathrm{FB}(P),
$$ 
which, together with $C_\mathrm{SK1}(P)=C_\mathrm{FB}(P)\ge C^g_\mathrm{SK2}(P)$, immediately implies \eqref{pr-equ-sup}. 

Finally, by Remark~\ref{remark-exact-Power:label}, for each $n$ there is $c_n\ge 1$ such that $(c_n\gamma_{1,n}, c_n\gamma_{2,n})$ is admissible and $P_2(c_n\gamma_{1,n}, c_n\gamma_{2,n}) = P$. Moreover, 
$$
\log r_n \leq R_2(c_n\gamma_{1,n}, c_n\gamma_{2,n})=\log r_n + 2\log c_n \leq C_\mathrm{FB}(P).
$$
Letting $n\to\infty$ and by \eqref{pf-ar1-F-rho-P} and \eqref{pf-ar1-lim-rP},  we obtain that
$$
\lim_{n\to\infty}R_2(c_n\gamma_{1,n}, c_n\gamma_{2,n})= C_\mathrm{FB}(P),
$$
as desired.
\end{proof}

Kim \cite[Section IV]{Kim2006} noted that an interleaving argument can show that Butman's original conjecture fails in general for $p>1$. We now consider an explicit family of stationary AR(2) Gaussian channels for which a genuine SK(2) coding scheme achieves feedback capacity and strictly outperforms SK(1).

\begin{pr}
\label{prop:interleaved-AR1}
Consider the stationary AR(2) Gaussian channel
$$
Y_i=X_i+Z_i,\qquad Z_i+\beta Z_{i-2}=W_i,\quad i\in\mathbb Z,
$$
where \(0<|\beta|<1\) and \(\{W_i\}\) is a zero-mean, unit-variance white Gaussian process. Then, for every $P>0$,
$$
C_{\mathrm{SK2}}(P)=C^g_{\mathrm{SK2}}(P) = C_{\mathrm{FB}}(P) > C_{\mathrm{SK1}}(P) = R_{\mathrm{Butman}}(P).
$$
Consequently, Conjecture~\ref{Butman-conj:label} is false for stationary AR(2) Gaussian channels.
\end{pr}

\begin{proof}
Define $Z_k^{(e)}=Z_{2k}$ and $Z^{(o)}_k=Z_{2k-1}$ for $k\in\mathbb{Z}$. Since the noise process $\{Z_i\}$ satisfies $Z_i+\beta Z_{i-2}=W_i$, we have
$$
Z^{(e)}_{k}+\beta Z_{k-1}^{(e)}=W_{2k}, \qquad Z^{(o)}_{k}+\beta Z^{(o)}_{k-1}=W_{2k-1}.
$$
Therefore, $\{Z^{(e)}_k\}$ and $\{Z^{(o)}_k\}$ are two independent stationary AR(1) Gaussian processes with parameter $\beta$, driven by the independent white Gaussian noises $\{W_{2k}\}$ and $\{W_{2k-1}\}$, respectively. Consequently, the original AR(2) Gaussian channel can be viewed as two independent parallel AR(1) Gaussian channels obtained by separating the even and odd channel uses.

Let $C_{\mathrm{FB}}(P;\beta)$ denote the feedback capacity of either AR(1) subchannel. Let $P_e$ and $P_o$ denote the average input powers per use of the even and odd AR(1) subchannels, respectively. Since each subchannel occupies one half of the original channel uses, the overall average-power constraint becomes
$$
\frac{P_e+P_o}{2}\le P.
$$
Consequently, the feedback capacity of the original channel is
\begin{equation*}\label{CFB-interleave-ar2}
\begin{aligned}
C_{\mathrm{FB}}(P)
&= \max_{\frac{P_{\mathrm e}+P_{\mathrm o}}{2}\le P}
\left\{
\frac12 C_{\mathrm{FB}}(P_{\mathrm e};\beta) + \frac12 C_{\mathrm{FB}}(P_{\mathrm o};\beta) \right\}.
\end{aligned}
\end{equation*}
Since $C_\mathrm{FB}(P;\beta)$ is concave and non-decreasing in $P$, we have
$$
\frac12 C_\mathrm{FB}(P_{\mathrm e};\beta) + \frac12 C_{\mathrm{FB}}(P_{\mathrm o};\beta)
\le
C_{\mathrm{FB}}\left(\frac{P_{\mathrm e}+P_{\mathrm o}}{2};\beta\right)
\le
C_{\mathrm{FB}}(P;\beta),
$$
with equality attained by $P_{\mathrm e}=P_{\mathrm o}=P$. Hence
\begin{equation}\label{CFB-interleave-ar2-2}
C_{\mathrm{FB}}(P)= C_{\mathrm{FB}}(P;\beta)=\log x_0,
\end{equation}
where $x_0>1$ is the unique solution of
\begin{equation}\label{eq:interleaved-capacity}
P= \frac{x_0^2-1} {\left(1+|\beta|x_0^{-1}\right)^2}.
\end{equation}

Now, applying the SK(2) coding scheme \eqref{SK2-coding-input}--\eqref{SK2-coding-message} and choosing the characteristic pair $(\gamma_1,\gamma_2)$ as
\begin{equation}\label{chara-pair-interleave-ar2}
(\gamma_1,\gamma_2) =
\begin{cases}
\left(\sqrt{x_0},-\sqrt{x_0}\right), & \beta>0,\\[1mm]
\left(\mathrm{i}\sqrt{x_0},-\mathrm{i}\sqrt{x_0}\right), & \beta<0.
\end{cases}
\end{equation}
Then
$$
\gamma_1+\gamma_2=0, \qquad \gamma_1\gamma_2=-\operatorname{sign}(\beta)x_0.
$$
In particular, 
$$
|\gamma_1|=|\gamma_2| = \sqrt{x_0}>1,
$$
and both $\gamma_1+\gamma_2$ and $\gamma_1\gamma_2$ are real. Moreover, 
$$
b = -\gamma_1\gamma_2=\operatorname{sign}(\beta)x_0\neq 0.
$$
Thus, it remains to verify that the characteristic pair $(\gamma_1,\gamma_2)$ satisfies the power constraint in Theorem~\ref{main-theorem-sk2}. For both $\beta>0$ and $\beta<0$, 
$$
\gamma_1^2=\gamma_2^2=\operatorname{sign}(\beta) x_0,
$$
and, since $L_Z(z)=1+\beta z^2$,
$$
L_Z(\gamma_1^{-1})=L_Z(\gamma_2^{-1}) = 1 +\frac{\beta}{\operatorname{sign}(\beta) x_0}=1 +\frac{|\beta|}{ x_0}.
$$
Moreover,
$$
c_1=c_2=\frac{1}{2\operatorname{sign}(\beta) x_0}.
$$
Substituting these expressions into the power constraint \eqref{power-eqn-sk2-thm} gives
$$
\frac{x_0^2-1}{(1 +|\beta| x^{-1}_0)^2}=P,
$$
where the equality follows from \eqref{eq:interleaved-capacity}. Thus, the chosen characteristic pair $(\gamma_1,\gamma_2)$ in \eqref{chara-pair-interleave-ar2} is feasible in Theorem~\ref{main-theorem-sk2}. Consequently,
$$
C^\mathrm{g}_\mathrm{SK2}(P)\ge \log|\gamma_1\gamma_2|=\log x_0.
$$
Since $C_\mathrm{SK2}(P)\ge C^\mathrm{g}_\mathrm{SK2}(P)$, together with \eqref{CFB-interleave-ar2-2}, it holds that
$$
 C_{\mathrm{FB}}(P)=\log x_0 \le C_{\mathrm{SK2}}(P) \le C_{\mathrm{FB}}(P),
$$
and hence
\begin{equation}\label{CSK2-CFB-AR2-Inter}
C^g_\mathrm{SK2}(P)=C_\mathrm{SK2}(P)=C_\mathrm{FB}(P) = \log x_0.
\end{equation}

It remains to prove that $C_{\mathrm{SK1}}(P)<C_{\mathrm{FB}}(P)$. To see this, Theorem~\ref{ISK1-thm} gives
\begin{equation}\label{CFB-sk1-inter}
C_\mathrm{SK1}(P)=\max\log |\gamma|,
\end{equation}
where the maximum is taken over all $\gamma\in\mathbb{R}$ with $|\gamma|>1$ satisfying 
\begin{equation}\label{eq:SK1-interleaved}
P= \frac{\gamma^2-1} {\left(1+\beta \gamma^{-2}\right)^2}.
\end{equation}
Let $\gamma_0$ be an optimizer of \eqref{CFB-sk1-inter} and define
$$
\Phi_\beta(x) = \frac{x^2-1}{\left(1+|\beta| x^{-1}\right)^2}, \qquad x>1.
$$
Since
$$
\Phi_\beta'(x) = \frac{2x\left(x^3+2|\beta| x^2-|\beta|\right)} {(x+|\beta|)^3} >0, \qquad x>1,
$$
the function $\Phi_\beta$ is strictly increasing. Moreover, 
$$
1+\frac{|\beta|}{|\gamma_0|} > 1+\frac{\beta}{\gamma_0^2} >0.
$$
Using \eqref{eq:SK1-interleaved}, we obtain
$$
\Phi_\beta(|\gamma_0|) < \frac{\gamma_0^2-1} {\left(1+\beta \gamma_0^{-2}\right)^2} = P.
$$
On the other hand, \eqref{eq:interleaved-capacity} gives
$$
\Phi_\beta(x_0)=P.
$$
The strict monotonicity of $\Phi_\beta$ implies $|\gamma_0|<x_0$. Consequently
$$
C_{\mathrm{SK1}}(P) = \log |\gamma_0| < \log x_0 = C_{\mathrm{FB}}(P),
$$
as desired.

Finally, it follows from $R_{\mathrm{Butman}}(P) = C_{\mathrm{SK1}}(P)$ and \eqref{CSK2-CFB-AR2-Inter} that 
$$
R_{\mathrm{Butman}}(P) < C_{\mathrm{SK2}}(P) = C_{\mathrm{FB}}(P),
$$
which completes the proof. 
\end{proof}

\subsection{Numerical Comparisons for AR(2) Channels}\label{subsec:numerical_analysis}
We next compare the performance of SK(1) and SK(2) numerically for stationary AR(2) Gaussian channels beyond the analytically tractable interleaved subclass. The examples illustrate strict SK(2) gains over intervals of the AR(2) parameters.

\begin{example}\label{example-ar2:label}
Consider a stationary AR(2) Gaussian channel with noise power spectral density
\begin{equation*}\label{AR2-SDF}
S_Z(e^{i\theta})=\frac{1}{|(1+\beta_1e^{i\theta})(1+\beta_2e^{i\theta})|^2},\qquad |\beta_1|,|\beta_2|<1.
\end{equation*}
This spectrum corresponds to the stationary noise process 
$$
Z_i +(\beta_1+\beta_2) Z_{i-1} +\beta_1\beta_2 Z_{i-2} = W_i,\quad i\in\mathbb{Z},
$$
where $\{W_i\}$ is a white Gaussian process with zero mean and unit variance. For the AR coefficients to be real, the pair $(\beta_1,\beta_2)$ must satisfy $\beta_1+\beta_2\in \mathbb{R}$ and $\beta_1\beta_2\in\mathbb{R}$ (i.e., $\beta_1$ and $\beta_2$ are either real or complex conjugates). 

Under average power constraint $P$, Theorem~\ref{ISK1-thm} yields $C_{\mathrm{SK1}}(P)= \max\log |x_0|$,
where the maximum is taken over all real $x_0$ with $|x_0|>1$ satisfying
\begin{equation*}\label{Power-Eqn-AR2}
P = \frac{x_0^2-1}{\left((1+ \beta_1 x_0^{-1})(1 + \beta_2 x_0^{-1})\right)^2}.
\end{equation*}
Numerical evaluation of the characterizations in Theorems~\ref{ISK1-thm} and \ref{main-theorem-sk2} gives the comparisons in Figure~\ref{fig:ISK-AR2-3Beta}. The computed curves indicate strict improvement over SK(1) on nontrivial parameter intervals.
By Corollary~\ref{main-col-sk2:label}, this implies
\begin{equation}\label{CISK2-ar2}
C_\mathrm{SK2}(P)= C^\mathrm{g}_\mathrm{SK2}(P)>C_{\mathrm{SK1}}(P)=R_\mathrm{Butman}(P).
\end{equation}
\begin{figure}[htbp!]
    \centering
    \centerline{\includegraphics[width=17cm]{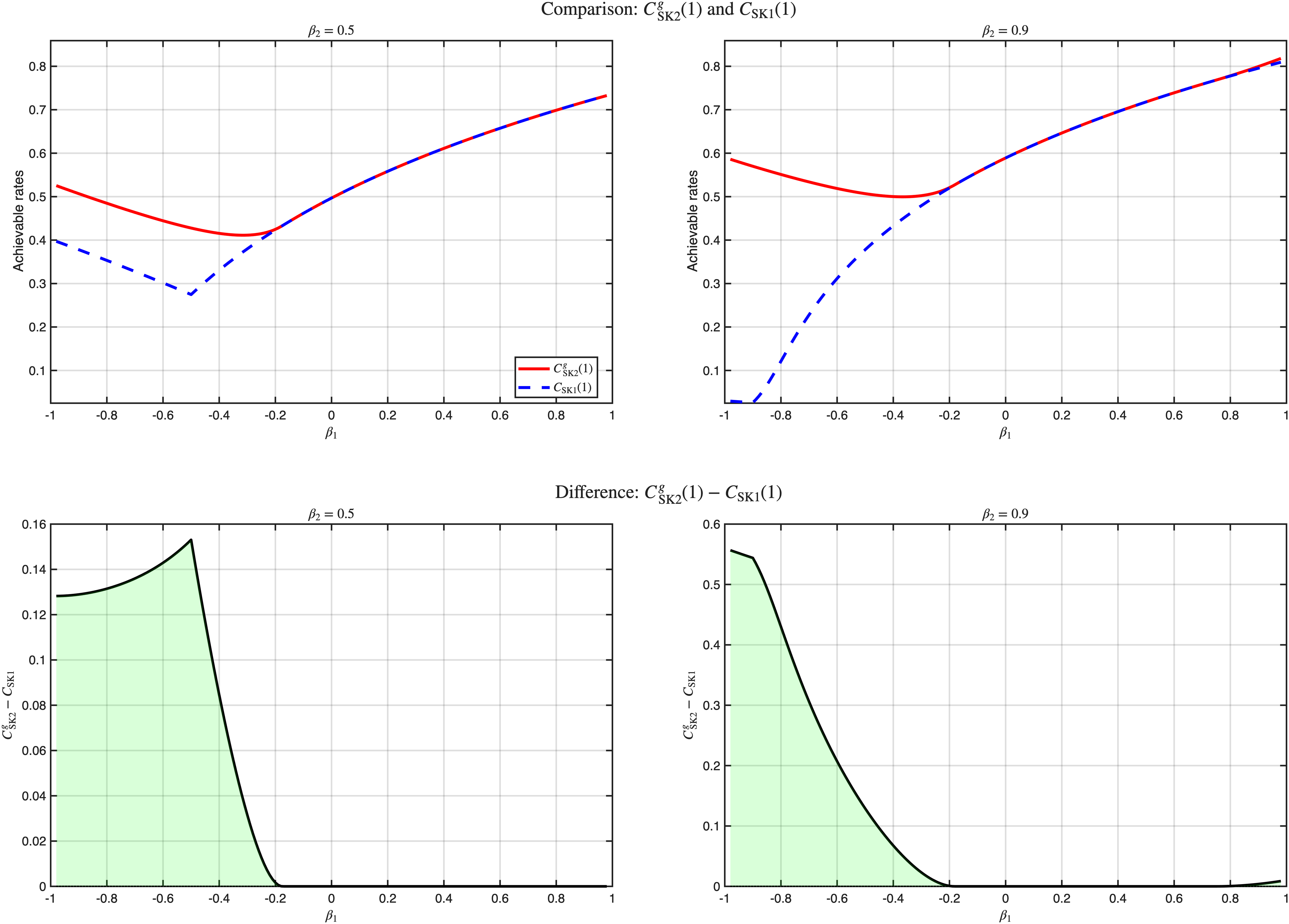}}
    \caption{Numerical comparison of SK(1) capacity and genuine SK(2) capacity for real AR(2) parameters \(\beta_1 \in (-1, 1)\) at \(P = 1\). The left and right columns correspond to \(\beta_2 = 0.5\) and \(\beta_2 = 0.9\), respectively. The upper panels plot \(C_{\mathrm{SK}1}(1)\) and the numerically evaluated \(C_{\mathrm{SK}2}^g(1)\); the lower panels plot their difference.}
    \label{fig:ISK-AR2-3Beta}
\end{figure}
\end{example}
Thus, the numerical results illustrate that the advantage of second-order message recursion occurs for a nontrivial range of AR(2) Gaussian channels.

\section{Conclusion and Outlook}

We introduced the SK(2) class of Gaussian feedback coding schemes for stationary AR($p$) Gaussian channels, in which the message process evolves according to a second-order deterministic recursion. We derived a closed-form characterization of its maximal achievable rate and showed that the SK(1) subclass provides a branch-complete and sign-consistent formulation of the real constant signed-ratio specialization of Butman's linear-feedback construction. For every stationary AR(\(p\)) Gaussian channel, we proved that
$$
C_{\mathrm{SK}2}(P) = C_{\mathrm{SK}2}^\mathrm{g}(P) \geq C_{\mathrm{SK}1}(P).
$$
Thus, restricting the optimization to recursions with two unstable characteristic roots does not change the rate supremum.

For the AWGN channel, fixed genuinely second-order recursions attain feedback capacity. For nonwhite AR(1) Gaussian channels, the genuine supremum equals feedback capacity but is not attained by any fixed admissible pair; the full SK(2) class attains capacity through its first-order specialization. For the interleaved AR(1) subclass of AR(2) Gaussian channels, an explicit genuinely second-order recursion attains feedback capacity and strictly outperforms SK(1), disproving the corrected constant signed-ratio form of Butman's conjecture stated in this paper. Numerical comparisons indicate further gains for other AR(2) parameters. These results identify the message-recursion order as a useful design parameter for Gaussian feedback coding.

General variational, algebraic, and state-space characterizations of feedback capacity are available \cite{kim2009feedback,liu2018feedback,sabag2023,fay2026feedback,su2026hardy}. Deriving tractable analytical formulas for the capacity and optimal message-recursion parameters directly from the noise coefficients remains an important problem. A natural direction is to investigate SK($q$) schemes with $q>2$, clarify their relationship to existing state-space constructions, and determine when a prescribed recursion order suffices to achieve feedback capacity. It is also of interest to determine whether increasing the recursion order yields a convergent hierarchy of achievable rates and how the noise spectrum determines the minimal effective message-recursion order.

\section*{Appendices}
\appendix

\section{Proof of Lemma \ref{lemma1}}\label{sec:proof_of_lemma_ref_lemma1}
For any nonzero $x\in \mathbb{R}^{n}$, since $S$ is symmetric positive definite, we have 
\begin{equation*}
\begin{aligned}
x^T (S + U U^T)x &= x^T S x + x^T U U^T x\\
&=x^T S x + \|U^T x  \|^2\\
&\geq x^T S x\\
&>0,
\end{aligned}
\end{equation*}
which shows that $S + U U^T$ is positive-definite and hence invertible. Then, we use the Woodbury matrix identity \cite{higham2002accuracy} to establish
\begin{equation*}
\begin{aligned}
(S +UU^T )^{-1} &= S^{-1} -S^{-1}U(I_m+U^T S^{-1} U)^{-1}U^T S^{-1}\\
&=S^{-1} - S^{-1}UA^{-1}U^T S^{-1}
\end{aligned}
\end{equation*}
and thus
\begin{equation*}
\begin{aligned}
U^T (S+U U^T)^{-1}U &= U^T S^{-1} U - U^T S^{-1} U A^{-1} U^T S^{-1}U\\
&=A-I_m -(A-I_m) A^{-1}(A-I_m)\\
&= I_m-A^{-1}.
\end{aligned}
\end{equation*}
This further immediately implies that for $1\le i,j\le m$
\begin{equation}\label{lemma-eqn-1}
u_i^T (S +UU^T)^{-1}u_j = \delta_{ij} - (A^{-1})_{ij},
\end{equation}
where $\delta_{ij}$ is the Kronecker delta function. Furthermore, since 
$$
A^{-1} = \frac{1}{\det(A)}\text{adj}(A),
$$
where \text{adj}$(A)$ is the adjugate matrix of $A$, it holds that 
$$
(A^{-1})_{ij} = (-1)^{i+j}\frac{1}{\det(A)} M_{ji},
$$
which, together with $(\ref{lemma-eqn-1})$, immediately completes this proof, as desired.

\section{Proof of Theorem \ref{ISK1-thm}}\label{sec:proof_of_theorem_ref_isk1_thm}
Consider the SK(1) coding scheme formulated by \eqref{SK1-coding-X} and \eqref{SK1-coding-message} over the stationary AR($p$) Gaussian channel \eqref{channel-arp} with the noise spectral density function $S_Z(e^{i\theta})$ given by \eqref{sdf-arp}. Fix $\gamma\in\mathbb{R}$ with $|\gamma|>1$. It follows directly from \eqref{SK1-coding-message} that $V_n=\gamma^{n-1}U$. Following the same arguments as used in \eqref{Y-tlta}--\eqref{d1d2}, we can express
\begin{equation}
   \widetilde{Y}_n = 
\begin{cases} 
\gamma^{n-1} U + Z_n, & n \le p, \\[2pt]
d_n U + W_n, & n \ge p+1,
\end{cases}
\end{equation}
where 
\begin{equation}\label{SK1-d}
   d_n = \gamma^{n-1}L_Z(\gamma^{-1}).
\end{equation}
Thus, the mutual information between $V_1^n$ and $Y_1^n$ can be expressed as 
\begin{equation}\label{mutual-info-rate-sk1}
   \begin{aligned}
      \frac{1}{n}I(V_1^n;Y_1^n)&= \frac{1}{n}I(U;Y_1^n)\\
      &=\frac{1}{n}I(U;\widetilde{Y}_1^n)\\
      &=-\frac{1}{2n}\log\E[|U-\E[U\mid\widetilde{Y}_1^n]|^2].
   \end{aligned}
\end{equation}
Let $G =  \begin{bmatrix}
         1 & \gamma &\ldots&\gamma^{p-1}
      \end{bmatrix}^T$, $D_n = \begin{bmatrix}
         d_{p+1} & d_{p+2}&\ldots&d_n
      \end{bmatrix}^T$ and $\Sigma_Z=\mathrm{Var}(Z_1^p)$. Since $U$ and $\widetilde{Y}_1^n$ are jointly Gaussian, it holds that
\begin{equation}\label{mmse-sk1-pf}
   \begin{aligned}
      \E[|U-\E[U\mid\widetilde{Y}_1^n]|^2] &= 1 - \begin{bmatrix}G^T &D_n^T \end{bmatrix}\begin{bmatrix}\Sigma_{Z}+GG^T&GD_n^T\\[2pt]D_nG^T&I+D_nD_n^T \end{bmatrix}^{-1}\begin{bmatrix}G \\[2pt]D_n\end{bmatrix}\\
      &=1 - \widetilde{D}_n^T\left(\begin{bmatrix}\Sigma_Z &0\\[2pt]0 &I\end{bmatrix}+\widetilde{D}_n\widetilde{D}_n^T\right)^{-1}\widetilde{D}_n \\
      &\overset{(a)}{=}  \frac{1}{1+G^T\Sigma_Z^{-1}G+D_n^TD_n}\\
      &= \frac{1}{1+G^T\Sigma_Z^{-1}G - L_Z^2(\gamma^{-1})G^TG +\frac{L_Z^{2}(\gamma^{-1})}{\gamma^{2}-1}(\gamma^{2n}-1)},
   \end{aligned}
\end{equation}
where  $\widetilde{D}_n = \begin{bmatrix} G^T &D_n^T
      \end{bmatrix}^T$ and $(a)$ follows from Lemma \ref{lemma1}.  Substituting \eqref{mmse-sk1-pf} into \eqref{mutual-info-rate-sk1} and taking the limit as $n\to\infty$, we obtain
\begin{equation}\label{proof-sk1-mr}
\begin{aligned}
\bar{I}(V;Y)=\lim_{n\to\infty}\frac{1}{n}I(V_1^n;Y_1^n) = \log|\gamma|.
\end{aligned}
\end{equation}
Moreover, following a similar approach as in \eqref{power-eqn-sk2}, we have
\begin{equation}\label{power-eqn-sk1-pf}
   \begin{aligned}
   \E[|X_n|^2] &=\E[|V_n - \E[V_n\mid Y_1^{n-1}]|^2]\\
            &= \E[|V_n - \E[V_n\mid \widetilde{Y}_1^{n-1}]|^2]\\
            &=\gamma^{2(n-1)}\E[|U- \E[U\mid\widetilde{Y}_{1}^{n-1}]|^2].
   \end{aligned}
\end{equation}
Substituting \eqref{mmse-sk1-pf} into \eqref{power-eqn-sk1-pf} and letting $n\to\infty$  yields
\begin{equation}\label{proof-sk1-power}
   \begin{aligned}
   \lim_{n\to\infty}\E[|X_n|^2] &= \frac{\gamma^2-1}{L_Z^2(\gamma^{-1})},
 \end{aligned}
\end{equation}
which, together with \eqref{power-constraint-sk1}, immediately implies that
$$
\lim_{n\to\infty} \frac{1}{n} \sum_{i=1}^n \mathbb{E}[X_i^2] = \frac{\gamma^2 - 1}{L^2_Z(\gamma^{-1})} \leq P.
$$
Define $P_1(\gamma)=(\gamma^2-1)/L^2_Z(\gamma^{-1})$ for $\gamma\in\mathbb{R}$ with $|\gamma|>1$. We have
\begin{equation}\label{eq:sup-sk1-rate}
C_\mathrm{SK1}(P)=\sup_\gamma \log |\gamma|,
\end{equation}
where the supremum is taken over all $\gamma\in\mathbb{R}$ with $|\gamma|>1$ satisfying $P_1(\gamma)\le P$. Note that the function $P_1(\gamma)$ extends continuously to $\gamma=\pm 1$ with value zero and tends to infinity as $|\gamma|\to\infty$. Hence, 
$$
\{ \gamma \in \mathbb{R} : |\gamma| \geq 1, \ P_1(\gamma) \leq P \}
$$
is compact and contains points with $|\gamma|>1$. The supremum in \eqref{eq:sup-sk1-rate} is therefore attained at some $\gamma_*$ satisfying $|\gamma_*| > 1$. If $P_1(\gamma_*) < P$, continuity would permit a small increase in \(|\gamma_*|\) while preserving feasibility, contradicting maximality. Thus, $P_1(\gamma_*) = P$, proving \eqref{I-SK1-thm}--\eqref{power-eqn-sk1-thm}. 


\section{Counterexamples for Butman's AR(p) Derivation}\label{sec:example_of_ar_2_gaussian_channel}
The following two examples illustrate the sign-compatibility and branch issues discussed in Section~\ref{subsec:relation_to_butman}.
\begin{example}\label{Butman-example-1:label}
Let $0<t<1$ and take
\[
    \beta_1=t,\qquad \beta_2=-t.
\]
Then
\[
    L_Z(z)=(1+tz)(1-tz)=1-t^2z^2.
\]
Hence
$$
    B(x)=-t^2x^{-2}<0,\qquad x\neq0 .
$$

For any \(P>0\), \eqref{butman-abs-eqn} becomes
\begin{equation}\label{Butman-Counterexample-pe-1}
    x^2 = 1+P\left(1+t^2x^{-2}\right)^2.
\end{equation}
Clearly, every real root of \eqref{Butman-Counterexample-pe-1} satisfies $|x|>1$. Let $r=|x|$. Under the additional constant signed-ratio restriction $x_i=x$, it follows from \eqref{Butman-ee-cond} that
\[
    s_i=\operatorname{sign}(x)^2s_{i-2}=s_{i-2}.
\]
On the other hand, since
\[
    S_i(r)=b_2s_{i-2}r^{-2}=-t^2s_{i-2}r^{-2},
\]
it holds that
\[
    \operatorname{sign}(S_i(r))=-s_{i-2}.
\]
Thus, the sign-selection rule \eqref{Butman-sign-cond} requires
\[
    s_i=-s_{i-2},
\]
which contradicts the equal-ratio requirement \(s_i=s_{i-2}\). This contradiction is independent of the sign of $x$. Therefore, no real root of \eqref{butman-abs-eqn} can be implemented together with both the sign-selection rule and the equal-ratio condition.

For completeness, the equation \eqref{Butman-Counterexample-pe-1} indeed has real roots. Setting \(u=x^2\), the equation \eqref{Butman-Counterexample-pe-1} becomes
\[
    u=1+P\left(1+\frac{t^2}{u}\right)^2 .
\]
Let
\[
    f(u)=u-1-P\left(1+\frac{t^2}{u}\right)^2.
\]

Then, $f(1)<0$, $\lim_{u\to\infty}f(u)=\infty$, and
\[
    f^\prime(u) = 1+\frac{2Pt^2}{u^2}\left(1+\frac{t^2}{u}\right)>0 .
\]
Consequently, there is a unique \(u_\ast>1\), and the real roots are \(x=\pm\sqrt{u_\ast}\). Both are incompatible with the recursive sign selection, as shown above.
\end{example}

\begin{example}
\label{Butman-example-2:label}
Let \(0<t<1\) and take $ \beta_1=ti,~\beta_2=-ti$.
Then
$$
    L_Z(z)=(1+tiz)(1-tiz)=1+t^2z^2.
$$
Consequently,
$$
    B(x)=t^2x^{-2}>0,\qquad x\in\mathbb R,\ x\neq0 .
$$
Hence,
\[
    |B(x)|=B(x)\neq -B(x)
\]
for every real \(x\neq0\). Therefore, the step from the absolute-value expression \eqref{butman-abs-eqn} to Butman's expression \eqref{butman-final-eqn-1} is not justified for this AR(2) Gaussian channel.
\end{example}

\section{Proof of (\ref{power-eqn-equivalence-case-c})}\label{sec:proof_of_ref_power_eqn_equivalence_case_c}
For notational simplicity, let 
\[
   f(\gamma_1,\gamma_2)= \frac{1}{\Delta}\left(\frac{L_{11}}{\gamma_2^2} + \frac{L_{22}}{\gamma_1^2} + \frac{2L_{12}}{\gamma_1\gamma_2}\right).
\]
It then follows that 
\begin{equation}\label{f}
   f(\gamma_1,\gamma_2) = \frac{(\gamma_1^2-1)(\gamma_2^2-1)(\gamma_1\gamma_2-1)^2}{(\gamma_1 - \gamma_2)^2L^2(\gamma_1^{-1})L^2(\gamma_2^{-1})}g(\gamma_1,\gamma_2),
\end{equation}
where 
\[
   g(\gamma_1,\gamma_2)=\frac{L^2(\gamma_1^{-1})}{\gamma_1^2-1} + \frac{L^2(\gamma_2^{-1})}{\gamma_2^2-1}-\frac{2L(\gamma_1^{-1})L(\gamma_2^{-1})}{\gamma_1\gamma_2-1}.
\]
Let $\gamma_2 = \gamma \in \mathbb{R}$. The Taylor expansion of $g(\gamma_1, \gamma)$ in $\gamma_1$ around $\gamma_1 = \gamma$ is:
\begin{equation}\label{taylor-g}
   \begin{aligned}
      g(\gamma_1,\gamma)= \sum_{n=0}^\infty\frac{1}{n!}\frac{\partial^ng}{\partial\gamma_1^n}(\gamma,\gamma)(\gamma_1-\gamma)^n
   \end{aligned}
\end{equation}
Clearly,
\begin{equation}\label{g-0}
   g(\gamma,\gamma)=0
\end{equation}
and
\begin{equation*}
   \begin{aligned}
      \frac{\partial g}{\partial \gamma_1}(\gamma_1,\gamma)= -\frac{2L(\gamma_1^{-1})L^\prime(\gamma_1^{-1})}{\gamma_1^2(\gamma_1^2-1)} -\frac{2\gamma_1L^2(\gamma_1^{-1})}{(\gamma_1^2-1)^2} + \frac{2L(\gamma^{-1})L^\prime(\gamma_1^{-1})}{\gamma_1^2(\gamma_1\gamma-1)}+\frac{2\gamma L(\gamma^{-1})L(\gamma_1^{-1})}{(\gamma_1\gamma-1)^2},
   \end{aligned}
\end{equation*}
which implies that 
\begin{equation}\label{g-1}
   \frac{\partial g}{\partial \gamma_1}(\gamma,\gamma)=0.
\end{equation}
The function $g(\gamma_1,\gamma_2)$ in \eqref{f} is analytic near every repeated admissible pair. Moreover, both \eqref{g-0} and \eqref{g-1} hold for every $\gamma_2$ in this neighborhood. Hence, $g(\gamma_1, \gamma_2) = (\gamma_1 - \gamma_2)^2 h_0(\gamma_1, \gamma_2)$ for a locally analytic function \(h_0\). Cancelling this factor in \eqref{f} gives a joint analytic extension across the diagonal. The fixed-coordinate calculation below identifies its diagonal value.
More specifically, we have 
\begin{equation*}
   \begin{aligned}
      \frac{\partial^2 g}{\partial \gamma^2_1}(\gamma_1,\gamma)&= \frac{2(L^\prime(\gamma_1^{-1}))^2}{\gamma_1^4(\gamma_1^2-1)} + \frac{2 L(\gamma_1^{-1}) L''(\gamma_1^{-1})}{\gamma_1^4 (\gamma_1^2 - 1)} - \frac{2L^2(\gamma_1^{-1})}{(\gamma_1^2 - 1)^2}+\frac{4(2\gamma_1^2-1)L(\gamma_1^{-1})L'(\gamma_1^{-1})}{\gamma_1^3(\gamma_1^2-1)^2}\\
      &~~~~ + \frac{4L(\gamma_1^{-1})L'(\gamma_1^{-1})}{\gamma_1(\gamma_1^2 - 1)^2}-\frac{2L(\gamma^{-1})L''(\gamma_1^{-1})}{\gamma_1^4(\gamma_1\gamma-1)}- \frac{2(3\gamma_1\gamma-2)L(\gamma^{-1})L'(\gamma_1^{-1})}{\gamma_1^3(\gamma_1\gamma-1)^2}\\
      &~~~~-\frac{2\gamma L(\gamma^{-1})L'(\gamma_1^{-1})}{\gamma_1^2(\gamma_1\gamma-1)^2} - \frac{4\gamma^2 L(\gamma^{-1})L(\gamma_1^{-1})}{(\gamma_1\gamma-1)^3}+\frac{8 \gamma_1^2 L^2(\gamma_1^{-1})}{(\gamma_1^2 - 1)^3},
   \end{aligned}
\end{equation*}
which implies that
\begin{equation}\label{g-2}
   \frac{\partial^2g}{\partial \gamma_1^2}(\gamma,\gamma)=\frac{2\left(L^\prime(\gamma^{-1})\right)^2}{\gamma^4(\gamma^2-1)} +\frac{4L(\gamma^{-1}) L'(\gamma^{-1})}{\gamma(\gamma^2-1)^2} +\frac{2(\gamma^2 + 1)\left(L(\gamma^{-1})\right)^2}{(\gamma^2 - 1)^3}.
\end{equation}
Therefore, combining \eqref{f}, \eqref{taylor-g}, \eqref{g-0}, \eqref{g-1} and \eqref{g-2} yields
\begin{equation*}
\begin{aligned}
   \lim_{\gamma_1 \to \gamma} f(\gamma_1, \gamma) &= \frac{(\gamma^2 - 1)^4}{2 L^4 (\gamma^{-1})} \frac{\partial^2 g}{\partial \gamma_1^2} (\gamma, \gamma)\\
   &=\frac{(\gamma^2 - 1)^3}{\gamma^4} 
\frac{\left(L'(\gamma^{-1})\right)^2}{L^4(\gamma^{-1})} + \frac{2 (\gamma^2 - 1)^2}{\gamma} \frac{L'(\gamma^{-1})}{L^3(\gamma^{-1})} + \frac{\gamma^4 - 1}{L^2(\gamma^{-1})},
\end{aligned}
\end{equation*}
from which (\ref{power-eqn-equivalence-case-c}) follows immediately.

\section{Proof of Lemma~\ref{lemma2:label}}\label{sec-app:reduction_lemma}
\begin{proof}
We follow the proof of Theorem~\ref{main-theorem-sk2} and retain its notation. The finite-dimensional calculations in \eqref{Y-tlta}--\eqref{MMSE-Vn2} do not require both roots to be unstable, so \eqref{pf-sk2-mutual-info} and \eqref{MMSE-Vn2} remain valid for all sufficiently large $n$.

Suppose first that $\gamma$ is the unique unstable root, and let $\delta$ be the other root. Since the characteristic polynomial has real coefficients, it holds that $\gamma,\delta\in\mathbb{R}$, $|\gamma|>1$, and $|\delta|\le 1$. It is easy to see that in this case
\begin{equation}\label{eq:one-unstable-an-bn}
a_n = \frac{\gamma\delta^{n-1} - \delta\gamma^{n-1}}{\gamma - \delta}, \qquad b_n = \frac{\gamma^{n-1} - \delta^{n-1}}{\gamma - \delta}.
\end{equation}
Moreover, $d_{1,n}$ and $d_{2,n}$ in \eqref{Y-tlta-2} are 
\begin{equation}\label{eq:one-unstable-d1-d2}
d_{1,n} = \frac{\gamma q_n - \delta L_Z(\gamma^{-1}) \gamma^{n-1}}{\gamma - \delta}, \qquad d_{2,n} = \frac{L_Z(\gamma^{-1}) \gamma^{n-1} - q_n}{\gamma - \delta}, \quad n \geq p + 1,
\end{equation}
where $L_Z(\gamma^{-1})\neq 0$ and $q_n\triangleq L_Z(B)\delta^{n-1}$ with the backward-shift operator $B$. In particular, when $\delta=0$, the sequence $\delta^{n-1}$ is interpreted as $(1, 0, 0, \ldots)$. If $0<|\delta|<1$, then $q_n=L_Z(\delta^{-1})\delta^{n-1}$ and $\{q_n\}_{n\ge p+1}$ is square summable; if $|\delta|=1$, then $q_n=L_Z(\delta)\delta^{n-1}$, where $L_Z(\delta)\neq 0$. It turns out from \eqref{eq:one-unstable-an-bn} and \eqref{eq:one-unstable-d1-d2} that
$$
a_n+\delta b_n=\delta^{n-1},\qquad d_{1,n}+\delta d_{2,n}=q_n,
$$
which, together with the same computations as in \eqref{2norm-D1n}--\eqref{det-sk2}, implies 
\begin{equation}\label{eq:det-lemma-3-4}
\det (\Delta_{2,n}) = 
\begin{cases} 
c\gamma^{2n} + o(|\gamma|^{2n}), & |\delta| < 1, \\[12pt]
\dfrac{L_Z^2(\gamma^{-1}) L_Z(\delta)^2}{(\gamma - \delta)^2(\gamma^2 - 1)} n\gamma^{2n} + O(|\gamma|^{2n}), & |\delta| = 1,
\end{cases}
\end{equation}
where $c$ is a positive constant independent of $n$. It then follows from \eqref{pf-sk2-mutual-info} and \eqref{eq:det-lemma-3-4} that
\begin{equation}\label{eq:mutual-lemma-3-4}
\bar{I}(V;Y) = \lim_{n\to\infty} \frac{1}{2n} \log \det (\Delta_{2,n}) = \log|\gamma|.
\end{equation}

We next evaluate \eqref{MMSE-Vn2}. Denote the numerator of \eqref{MMSE-Vn2} by $N_n$. Substituting \eqref{eq:one-unstable-an-bn} and \eqref{eq:one-unstable-d1-d2} into $N_n$ and collecting the leading terms, as in \eqref{aD-bD}--\eqref{power-eqn-sk2-proof}, we obtain
\begin{equation}\label{eq:Nn-lemma-3-4}
    N_n = 
\begin{cases} 
c\dfrac{\gamma^2 - 1}{L_Z^2(\gamma^{-1})}\gamma^{2n-2} + o(|\gamma|^{2n-2}), & |\delta| < 1, \\[12pt]
\dfrac{L_Z(\delta)^2}{(\gamma - \delta)^2} n\gamma^{2n-2} + O(|\gamma|^{2n-2}), & |\delta| = 1,
\end{cases}
\end{equation}
where $c$ is the same constant as in \eqref{eq:det-lemma-3-4}. Therefore, combining \eqref{eq:det-lemma-3-4} and \eqref{eq:Nn-lemma-3-4} with \eqref{MMSE-Vn2}, with $n-1$ in the denominator, yields
\begin{equation}
    \lim_{n\to\infty}\mathbb{E}[X_n^2] = \frac{\gamma^2 - 1}{L_Z(\gamma^{-1})^2},
\end{equation}
which, together with \eqref{power-constraint-sk1}, immediately implies that
\begin{equation}\label{eq:power-lemma-3-4}
\lim_{n\to\infty} \frac{1}{n} \sum_{i=1}^n \mathbb{E}[X_i^2]=\frac{\gamma^2 - 1}{L_Z(\gamma^{-1})^2}   \leq P.
\end{equation}
Consequently, combining \eqref{eq:mutual-lemma-3-4}, \eqref{eq:power-lemma-3-4} and Theorem~\ref{ISK1-thm}, we obtain that 
\begin{equation}\label{eq:mutual-1-unstable-root}
    \bar{I}(V;Y)=\log |\gamma|\le C_\mathrm{SK1}(P).
\end{equation}

We next consider the case when neither root is unstable. In this case, it follows from \eqref{SK2-coding-message}, \eqref{Vn-case-a} and \eqref{Vn-case-b} that 
$$
|a_n|+|b_n|=O(1+n),\qquad |d_{1,n}| + |d_{2,n}|=O(1+n).
$$
As a result, by \eqref{pf-sk2-mutual-info}, we have $\mathrm{tr}(\Delta_{2,n})=O(1+n^3)$. By the definition of $\Delta_{2,n}$ in \eqref{pf-sk2-mutual-info}, we have $\Delta_{2,n} \succeq I_2$ and then
$$
1 \leq \det (\Delta_{2,n}) \leq \left( \frac{\operatorname{tr} (\Delta_{2,n})}{2} \right)^2.
$$
Therefore,
$$
0\le\frac{1}{n} I(V_1^n;Y_1^n)=\frac{1}{2n}\log \det (\Delta_{2,n})\le \frac{1}{n}\log\frac{\operatorname{tr} (\Delta_{2,n})}{2} = O\left(\frac{\log(n + 1)}{n}\right).
$$
It follows that
\begin{equation}\label{eq:mutual-0-unstable-root}
    \bar{I}(V;Y)=0.
\end{equation}
Thus, the asserted upper bound holds for every power-feasible recursion with no unstable characteristic root. Combining \eqref{eq:mutual-1-unstable-root} and \eqref{eq:mutual-0-unstable-root} completes the proof.
\end{proof}

\bibliographystyle{ieeetr}
\bibliography{SK2_TIT}

\end{document}